\documentclass[aps, reprint, prl, twocolumn, amsmath, amssymb, groupedaddress, 10pt, longbibliography, superscriptaddress]{revtex4-1}

\usepackage{epsfig}
\usepackage{subfigure}
\usepackage{graphicx}
\usepackage{mwe}
\usepackage{duckuments}
\usepackage{amsfonts}
\usepackage[figuresright]{rotating}
\usepackage{amssymb}
\usepackage{amsmath}
\usepackage{psfrag}
\usepackage{esint} \usepackage{hyperref}
\hypersetup{colorlinks,allcolors=blue} \usepackage{multirow}
\usepackage{siunitx}
\usepackage{xcolor}
\usepackage{lipsum}
\usepackage{bbm}
\usepackage[normalem]{ulem}

\usepackage[T5]{fontenc}

\begin{document}

\title{
Smectic Superconductivity }

\author{Grayson R. Frazier}
\affiliation{Department of Physics and Astronomy, Johns Hopkins University, Baltimore, Maryland 21218, USA}
\affiliation{Kavli Institute for Theoretical Physics, University of California, Santa Barbara, CA 93106, USA}
\author{Erez Berg}
\affiliation{Department of Condensed Matter Physics, Weizmann Institute of Science, Rehovot 76100, Israel}
\affiliation{Materials Department, University of California Santa Barbara, Santa Barbara 93106 USA}
\affiliation{Department of Electrical and Computer Engineering, University of California, Santa Barbara, CA 93106, USA}

\begin{abstract}
    We identify a mechanism in which soft nonunitary fluctuations of the pair condensate can suppress phase coherence in a two-dimensional Rashba superconductor proximate to easy-plane ferromagnetic order.
This spin polarization couples linearly to the supercurrent through a Lifshitz invariant, acting as an emergent gauge field for the superconducting $\mathrm{U}(1)$ phase.
    Integrating out soft magnetic fluctuations renormalizes the effective superfluid stiffness and drives the system into a ``smectic'' superconducting state, characterized by a finite-momentum helical phase texture and a long-wavelength stiffness that vanishes in one direction.
    Consequently, vortices no longer have a logarithmically divergent energy cost, and the quasi-long-range superconducting order is destroyed at any finite temperature.
    Weak lattice anisotropy 
pins the ferromagnetic order, cutting off the smectic-like behaviour at long wavelengths and thereby restoring a finite effective stiffness.
    Nonetheless, near the onset of magnetic order, the enhanced magnetic susceptibility leads to a vortex-deconfined regime for arbitrarily weak magnetoelectric coupling.
    The results provide insight into how intertwined superconducting and easy-plane ferromagnetic orders can reshape phase coherence, with relevance to rhombohedral graphene proximitized by a transition-metal dichalcogenide substrate.
\end{abstract}

\date{September 17, 2026}

\maketitle

Superconductivity is determined not only by the local pairing amplitude but also by the $\mathrm{U}(1)$ phase stiffness.
For the standard $XY$-model in two dimensions, the superfluid stiffness governs the Berezinskii-Kosterlitz-Thouless (BKT) transition, below which vortices are logarithmically confined and there is quasi-long-range order (QLRO)~\cite{Berezinskii1971, Berezinskii1972, Kosterlitz1973, Nelson1977, Minnhagen1987}.
This description typically assumes that the superconducting $\mathrm{U}(1)$ phase is the only relevant soft degree of freedom.
However, in unconventional superconductors, the pairing order can carry additional internal degrees of freedom, such as spin or orbital structure~\cite{Sigrist1991, Mineev1999}.
Variations in the internal pairing structure can couple directly to the superconducting $\mathrm{U}(1)$ phase gradient through, for example, Lifshitz invariants or emergent gauge fields, thereby driving nonuniform superconducting textures~\cite{Mermin1976, Mineev1994a, Mineev2008, Volovik2009, Agterberg2012, Frazier2026a}.
An open question is how fluctuations in the internal pairing structure can qualitatively change the superconducting state and affect phase coherence, even when the local pairing amplitude persists.

Rhombohedral graphene, consisting of $ABC$-stacked graphene, provides a natural platform for exploring this interplay.
The system features a rich family of correlated phases~\cite{Zhou2021a, Seiler2022, Barrera2022, Liu2023, Han2023, Lu2024, Xie2025} and also unconventional superconductivity~\cite{Chen2019, Zhou2021, Zhou2022, Choi2025, Han2025, Kumar2026, Yang2026}.
When proximitized by a transition-metal dichalcogenide (TMD) substrate, the rhombohedral graphene acquires enhanced spin-orbit coupling, paving way for possible coexistence or interaction between superconductivity and nearby spin-polarized and isospin-ordered phases~\cite{Zhang2023, Yang2025, Patterson2025, Holleis2025, Seo2026, Zhang2026, Qin2026, Kumar2026a, Deng2026}.
The vicinity of superconductivity to such symmetry-broken phases has motivated proposals exploring how collective fluctuations mediate unconventional pairing in
rhombohedral graphene~\cite{Chatterjee2022a, Dong2023, Dong2026, Mayrhofer2026, Raines2026, Calvera2026}.

Here, we seek to understand how such fluctuations can instead reshape the superconducting phase itself, modifying the phase stiffness, vortex energetics, and BKT phenomenology.
We demonstrate that fluctuations in easy-plane ferromagnetic order can suppress phase coherence in a spin-orbit-coupled superconductor.
In the presence of, for example, Rashba spin-orbit coupling, the system admits a Lifshitz invariant in which the superconducting $\mathrm{U}(1)$ phase gradients couple linearly to the in-plane spin polarization.
Soft nonunitary fluctuations of the spin-polarized pairing act as an emergent gauge field and screen the effective phase stiffness of the superconducting condensate, driving a smectic-like phase with vanishing long-wavelength stiffness.
This screening leads to a vortex-deconfined region near the onset of ferromagnetic order, and QLRO can be restored by in-plane pinning of the spin polarization.
We discuss the implications of the magnetoelectric coupling on the phase diagram and the relevance to rhombohedral graphene in proximity to transition-metal dichalcogenides.

 \phantomsection
\addcontentsline{toc}{section}{Nonunitary fluctuations as an emergent gauge field}
\paragraph*{{Nonunitary fluctuations as an emergent gauge field.---}}
We consider a minimal model of a two-dimensional superconductor coexisting with weakly pinned easy-plane ferromagnetic order.
In the absence of inversion symmetry, spin singlet and triplet pairing correlations are generally mixed~\cite{Sigrist1991, Gorkov2001, Agterberg2012}, with the superconducting order parameter of the form
\begin{math}
    \hat{\Delta}_{\alpha \beta}(\mathbf{r}, \mathbf{k})
    =
    e^{i\theta(\mathbf{r})}
    [(
    \psi(\mathbf{r}, \mathbf{k})
    +
    \mathbf{d}(\mathbf{r}, \mathbf{k}) \cdot \boldsymbol{\sigma}
    )i  \sigma^y]_{\alpha \beta}.
\end{math}
Here, $\alpha, \beta  = \uparrow, \downarrow$ are spin indices, $\theta(\mathbf{r})$ is the $\mathrm{U}(1)$ superconducting phase, $\sigma^i$ are Pauli spin matrices, $\mathbf{r}$ is the center-of-mass coordinate of the Cooper pair, and $\mathbf{k}$ is the relative momentum.
The field $\psi(\mathbf{r}, \mathbf{k}) = \psi(\mathbf{r}, -\mathbf{k})$ describes spin singlet component, and $\mathbf{d}(\mathbf{r}, \mathbf{k}) = -\mathbf{d}(\mathbf{r}, -\mathbf{k})$ is the $d$ vector describing spin triplet component~\cite{Balian1963, Leggett1975, Mackenzie2003, Volovik2009}.
We focus on the nonunitary spin polarization of the condensate,
\begin{math}
    \mathbf{m}(\mathbf{r})
    =
    (1/ 4 \pi^2)
    \int \mathrm{d}^2 k \,
    i \mathbf{d}(\mathbf{r}, \mathbf{k}) \times \mathbf{d}^*(\mathbf{r}, \mathbf{k}),
\end{math}
which measures the local nonunitarity of the triplet pairing state.
In the coexistence phase, $\mathbf{m}$ is locked to the easy-plane magnetization.
The singlet component, while allowed by symmetry, does not affect the form of the symmetry-allowed coupling discussed below and rather may be treated as renormalizing the phenomenological coefficients
in the effective theory below.

The minimal free energy density is given by
\begin{align}
    \nonumber
    f[\mathbf{q}(\mathbf{r}), \mathbf{m}(\mathbf{r})]
    &=
    \frac{\rho_s}{2}
    (\mathbf{q} - \lambda \hat{\mathbf{z}} \times \mathbf{m})^2
    +
    \frac{K_m}{2} (\partial_i \mathbf{m}) \cdot (\partial_i \mathbf{m})
    \\
    &\hspace{1em}
    +
    \frac{r_m}{2} \mathbf{m}^2
    +
    \frac{u_m}{4} (\mathbf{m}^2)^2,
\label{free_energy_density}
\end{align}
with summation implied over repeated indices $i = x, y$.
Above, $\rho_s$ is the bare charge stiffness, $K_m$ is the stiffness of the spin texture, and $r_m$ and $u_m$ are phenomenological Landau coefficients.
The gauge-invariant superconducting phase gradient is given by $\mathbf{q} (\mathbf{r}) = \boldsymbol{\nabla} \theta(\mathbf{r}) - \frac{2\pi}{\Phi_0} \mathbf{A}_\mathrm{EM}(\mathbf{r})$, with $\Phi_0 = h/2e$ being the flux quantum of a Cooper pair and $\mathbf{A}_\mathrm{EM}$ the electromagnetic vector potential.

The first term in the free energy encodes the magnetoelectric coupling allowed by broken inversion symmetry, corresponding to a Lifshitz invariant~\cite{Edelstein1995, Edelstein1996, Smidman2017, Kapustin2022, Levitan2025},
\begin{equation}
    f_\mathrm{ME} = \lambda \rho_s \hat{\mathbf{z}} \cdot (\mathbf{q} \times \mathbf{m}).
    \label{ME_coupling}
\end{equation}
Equivalently, the spin polarization acts as an emergent gauge field felt by the superconducting phase,
\begin{math}
    \boldsymbol{\mathcal{A}} (\mathbf{r})
    =
    \lambda \hat{\mathbf{z}} \times \mathbf{m} (\mathbf{r}),
\end{math}
and
the condensate lowers its energy by aligning its phase gradient with $\boldsymbol{\mathcal{A}}(\mathbf{r})$.
In the presence of uniform spin polarization, the pairing order is a Fulde-Ferrell-like state, having a helical texture with a spatially varying $\mathrm{U}(1)$ phase~\cite{Fulde1964, Larkin1965, Mineev1994a, Agterberg2003, Agterberg2020}.
Such a term can arise from, for example, Rashba spin-orbit coupling in the presence of an exchange field~\cite{Dimitrova2007, Mineev2008, SM}.

 \phantomsection
\addcontentsline{toc}{section}{Smectic superconductivity 
and vortex deconfinement}
\paragraph*{{Smectic superconductivity 
and vortex deconfinement.---
}} We now expand about the magnetized saddle point.
Although a rotationally symmetric easy-plane magnet possesses no true long-range order at finite temperature in two dimensions, its amplitude is locally well-defined, and we may expand about a locally ordered configuration.
For $r_m <0$, 
we take 
$\mathbf{m}_0 = m_0 \hat{\mathbf{x}}$ without loss of generality, in which $m_0^2 = -r_m/u_m$.
At the uniform saddle point, the supercurrent vanishes, with the superconducting phase gradient following the emergent gauge field, $\mathbf{q}_0 = \lambda \hat{\mathbf{z}} \times \mathbf{m}_0 = \lambda m_0 \hat{\mathbf{y}}$.
We write small fluctuations about this saddle point as $\mathbf{m} = \mathbf{m}_0 + \delta \mathbf{m}$ and $\mathbf{q} = \mathbf{q}_0 + \delta \mathbf{q}$, with the effective quadratic free energy density given by
\begin{align}
    \nonumber
    f^{(2)}[\delta\mathbf{q}, \delta \mathbf{m}]
    &=
    \frac{\rho_s}{2} (\delta q_x + \lambda \delta m_y)^2
    +
    \frac{\rho_s}{2} (\delta q_y - \lambda \delta m_x)^2
    \\
    &\hspace{1em}
    +
    \frac{K_m}{2}(\partial_i \delta \mathbf{m})
    \cdot
    (\partial_i \delta \mathbf{m})
    +
    \frac{\mu_\parallel}{2} (\delta m_x)^2.
\end{align}
Here, $\mu_\parallel = -2 r_m$ is the 
longitudinal 
mass of the ordered magnetic state.
The longitudinal fluctuation $\delta m_x$ is a gapped amplitude mode of the ferromagnetic order, while $\delta m_y$ corresponds to the transverse in-plane fluctuation, which is gapless without the presence of additional pinning terms.

Notably, a uniform phase twist $\delta q_x$ can be absorbed by a uniform rotation of the magnetization, $\delta m_y = - \delta q_x/ \lambda$.
When the transverse mode is gapless, this compensation costs no energy, and the long-wavelength phase stiffness along $\hat{\mathbf{x}}$ is eliminated.
Integrating out Gaussian fluctuations $\delta \mathbf{m}$ yields the phase-only effective free energy~\cite{SM},
\begin{equation}
    F_\mathrm{eff}[\delta \mathbf{q}]
    =
    \frac{1}{2}
    \int \frac{\mathrm{d}^2 k}{(2\pi)^2}
    \Big[
    \rho_\mathrm{eff}^x (\mathbf{k}) |\delta q_x(\mathbf{k})|^2 + \rho_\mathrm{eff}^y(\mathbf{k}) |\delta q_y (\mathbf{k})|^2
    \Big],
\end{equation}
in which the
effective anisotropic phase stiffness is given by
\begin{subequations}
    \begin{align}
        \rho_\mathrm{eff}^x(\mathbf{k})
        &=
        \frac{\rho_s K_m k^2}{K_m k^2 + \rho_s \lambda^2},
        \\
        \rho_\mathrm{eff}^y(\mathbf{k}) 
        &= 
        \frac{\rho_s(K_m k^2 + \mu_\parallel)}{K_mk^2 + \mu_\parallel + \rho_s \lambda^2}.
    \end{align}
    \label{effective_stiffness}\end{subequations}
In the ordered phase, the longitudinal mode $\delta m_x$ is massive.
For weak magnetoelectric coupling, 
$\mu_\parallel \gg \rho_s \lambda^2$,
the long-wavelength stiffness in the $y$ direction is given by the bare superfluid stiffness, $\rho_\mathrm{eff}^y(0) \approx \rho_s$.
In contrast, soft transverse nonunitary fluctuations screen the superfluid stiffness along the $x$ direction, which vanishes for long wavelengths, $\rho_\mathrm{eff}^x(\mathbf{k}) \approx K_m k^2/\lambda^2$.
A related suppression of the superfluid stiffness also occurs near the field-induced Lifshitz point in Rashba superconductors~\cite{Dimitrova2007}; here, however, the reduced superfluid stiffness remains suppressed throughout the unpinned coexistence phase, owing to the gapless transverse magnetic mode.

\begin{figure}
    \centering
    \includegraphics[width=\linewidth]{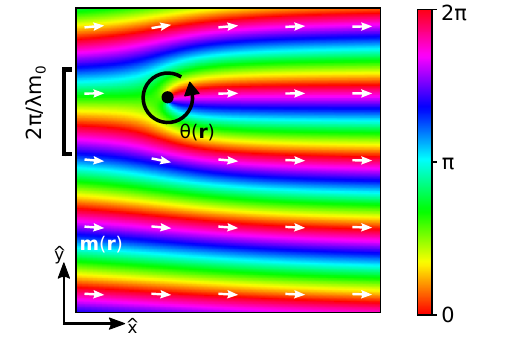}
    \caption{
    Illustration of a vortex in a smectic superconductor. The color represents the superconducting phase. 
    Through the magnetoelectric coupling, the spin polarization of the condensate acts as an emergent gauge field for the superconducting phase.
    The $\mathrm{U}(1)$ phase gradient is locked to the transverse nonunitary magnetization, displaying a helical texture with pitch $2\pi/|\lambda m_0|$.
    In absence of in-plane pinning of the transverse fluctuations, a vortex defect (black) has finite energy cost and QLRO is destroyed.
    White arrows indicate the local in-plane magnetization, $\mathbf{m}(\mathbf{r})$.
}
    \label{fig:smectic}
\end{figure}

Upon integrating out the fluctuations in the spin polarization, the effective theory maps to that of a smectic~\cite{Gennes1972, Gennes1993, Chaikin1995}.
Namely,
for smooth phase fluctuations $\delta q_x = ik_x \delta \theta$ and $\delta q_y = ik_y \delta \theta$, the effective free energy is given by
\begin{equation}
    F_\mathrm{eff} [\delta \theta]
    \approx 
\frac{1}{2} \int \mathrm{d}^2r
    \left[
    \rho_s (\partial_y \delta \theta)^2 + \frac{K_m}{\lambda^2} (\boldsymbol{\nabla} \partial_x \delta \theta)^2
    \right].
    \label{smectic_superconductor.free_energy_of_phase_fluctuations}
\end{equation}
This has the form of a smectic elastic energy, with $\delta \theta$ playing the role of the displacement of the density modulation. 
The stiffness $\rho_s$ corresponds to the compression modulus along the layer normal $\hat{\mathbf{y}}$, and the second term is a bending modulus that stabilizes distortions along the in-layer direction $\hat{\mathbf{x}}$.
The helical saddle point spontaneously selects a direction and produces a spatially modulated phase texture, as shown in Fig.~\ref{fig:smectic}.
Within this analogy, the magnetization plays the role of the nematic director: through the magnetoelectric coupling, it acts as an emergent gauge field for the $\mathrm{U}(1)$ phase.
Whereas the gauge fluctuations in the smectic-$A$ analogy render the transition weakly first order~\cite{Halperin1974}, here, the soft transverse fluctuations instead collapse the long-wavelength phase stiffness along one direction, leaving the local pairing amplitude finite. 
Related smectic-like phase elasticity has been discussed in finite-momentum atomic superfluids, including $p$-wave atomic-molecular condensates~\cite{Radzihovsky2009, Choi2011} and frustrated helical Bose condensates~\cite{Hsieh2022}.
Here, in contrast, the smectic behaviour emerges in a solid-state system through magnetoelectric coupling.

Due to the screened stiffness along one direction, the BKT transition temperature is suppressed.
Consider a vortex excitation satisfying
\begin{math}
    \boldsymbol{\nabla} \times \boldsymbol{\nabla} \delta \theta = 2\pi \delta(\mathbf{r}),
\end{math}
as shown in Fig.~\ref{fig:smectic}.
In contrast to typical $XY$ systems with logarithmically diverging vortex cost, the vortex cost in the smectic system is finite~\cite{SM},
\begin{equation}
    F_\mathrm{vortex} 
=
    \frac{\pi\sqrt{K_m \rho_s}}{|\lambda|}
    \Lambda,
\end{equation}
with $\Lambda$ being a momentum cutoff.
Consequently, 
the BKT temperature is driven to $T_\mathrm{BKT} \rightarrow 0$, with the QLRO destroyed at finite temperature.
The finite vortex energy is the superconducting analogue of dislocation deconfinement in a two-dimensional smectic~\cite{Toner1981, Gennes1993}.
Here, in contrast, the smectic elasticity is not imposed phenomenologically but rather emerges from the magnetoelectric coupling between soft nonunitary fluctuations and the $\mathrm{U}(1)$ phase gradient in Eq.~\eqref{ME_coupling}.
As such, for weak pinning, the state corresponds to that of a ``smectic superconductor,'' with a helical $\mathrm{U}(1)$ phase and vortex deconfinement at finite temperature.
The analogy of an FFLO superconductor to a smectic has been pointed out previously in the context of stripe and pair-density-wave order, where unidirectional modulation of the charge or pairing amplitude breaks translational and rotational symmetry~\cite{Agterberg2020}. 
Smectic-like electronic phases have been proposed in Refs.~\citenum{Kivelson1998} and \citenum{Emery2000}. 
Here, the pairing amplitude is spatially uniform, with the smectic behaviour emerging from the effective elasticity of the superconducting $\mathrm{U}(1)$ phase.

The magnetoelectric coupling has relevance to superconducting rhombohedral graphene proximitized by TMDs.
The substrate induces both Ising and Rashba spin-orbit coupling~\cite{Gmitra2015, Gmitra2016},
and local magnetic measurements suggest a region of coexistence between superconductivity and a spin-canted ferromagnetic order~\cite{Patterson2025, Zhang2026}.
In the coexistence region, Rashba spin-orbit coupling permits the Lifshitz invariant in Eq.~\eqref{ME_coupling}.
In contrast, Ising valley-spin locking alone retains the in-plane spin-rotation symmetry and is insufficient to realize the magnetoelectric coupling.
The combination of Rashba spin-orbit coupling and exchange coupling to the in-plane magnetization
yields a Lifshitz coefficient that is approximately linear in the exchange coupling and Rashba spin-orbit coupling~\cite{SM}.
First-principles calculations and experimental measurements on graphene on TMDs typically predict a proximity-induced Ising spin-orbit coupling of up to about $2$~meV, whereas the strength of the Rashba spin-orbit coupling has reported values varying widely from $0.1$ to $15$~meV~\cite{Gmitra2015, Wang2016, Gmitra2016, Gmitra2017, Wang2019, Island2019, Fueloep2021, Sun2023a, Amann2022, Masseroni2024}.
For conservative estimates of the Rashba spin-orbit coupling on the order of $1$~meV with bare superfluid stiffness of about $5$~meV, we find $|\lambda m_0|$ to be on the order of $10^{-5}$--$10^{-6}$~nm$^{-1}$, corresponding to a helical pitch on the order of $0.1$--$1$~mm.
For stronger Rashba SOC on the order of $10$~meV with a reduced bare superfluid stiffness on order of $1$ meV, this can increase the coupling to the order of $10^{-4}$--$10^{-3}$~nm$^{-1}$, yielding pitches on the order of one to tens of microns.
Notably, the smectic-like behaviour does not require a large Rashba spin-orbit coupling.
Namely, in absence of magnetic pinning, any finite $\lambda$ leads to a smectic-like regime, with the long-wavelength stiffness vanishing.
Though even a small microscopic magnetoelectric coupling can qualitatively reshape the finite-temperature phase diagram, 
the effect may be difficult to observe when the system size is smaller than the helical pitch, as we review in the discussion section.

 \phantomsection
\addcontentsline{toc}{section}{In-plane pinning and restoration of quasi-long-range order}
\paragraph*{{In-plane pinning and restoration of quasi-long-range order.---}}

Although transverse nonunitary fluctuations can screen the phase stiffness, pinning of the ferromagnetic order due to crystalline anisotropy can introduce a logarithmic vortex interaction at long length scales, thereby restoring a finite BKT transition temperature.
Motivated by rhombohedral graphene, we focus on the crystalline pinning in a system with $C_3$ symmetry and time-reversal symmetry.

For a fixed spin polarization amplitude,
\begin{math}
    \mathbf{m}(\mathbf{r}) = m_0 (\cos \varphi(\mathbf{r}), \sin \varphi(\mathbf{r}))^\mathrm{T},
\end{math}
the combined $C_3$ rotation and time reversal symmetries give rise to a sixfold anisotropy.
The long-wavelength free energy for angle $\varphi$ is given by
\begin{equation}
    F_\varphi[\varphi(\mathbf{r})]
    =
    \int \mathrm{d}^2 r
    \left(
    \frac{\kappa}{2} (\boldsymbol{\nabla} \varphi)^2 - g \cos (6 \varphi)
    \right),
\end{equation}
in which $\kappa = K_m m_0^2$ and $g$ parameterizes the bare sixfold clock anisotropy.
For $g=0$, the Gaussian free-field correlator is given by
\begin{math}
    \langle e^{i 6 \varphi(\mathbf{r})} e^{-i 6 \varphi(0)} \rangle
    \propto
    r^{-36 T/(2\pi \kappa)},
\end{math}
yielding scaling dimension
\begin{math}
    d_6 = {9 T}/{\pi \kappa}.
\end{math}
Close to the transition in which the $Z_6$ clock symmetry is broken, the system obeys the BKT-type flow~\cite{Jose1977, Gali2024},
${\mathrm{d} y}/{\mathrm{d} \ell} = xy$ and ${\mathrm{d} x}/{\mathrm{d} \ell} = Ay^2$, 
in which $A > 0$, $x = 2 - d_6$, and $y$ is the dimensionless fugacity.
The sixfold anisotropy is relevant for $d_6 < 2$, or equivalently, for
\begin{math}
    {\kappa}/{T} = {K_m m_0^2}/{T} > {9}/{2\pi}.
\end{math}
The lower transition into the $Z_6$ ordered phase is given by $T_{\mu_\perp} \approx 2 \pi \kappa/9$.
For $T < T_{\mu_\perp}$, the sixfold anisotropy flows to strong coupling and locks the in-plane angle $\varphi$ to one of six preferred orientations,
leading to a long-range-ordered ferromagnetic phase.

Near the transition $T_{\mu_\perp}$, the corresponding crossover scale has the BKT form
\begin{math}
    \xi_{\mu_\perp} 
\sim a \exp
    (
    {b}/{\sqrt{(T_{\mu_\perp} - T)/T_{\mu_\perp}}}
    ),
\end{math}
in which $a$ is a short-distance cutoff and $b$ is a phenomenological constant of order unity.
At scales $r \gg \xi_{\mu_\perp}$,
the orientation angle is locked to one of the six orientations.
Expanding about a chosen direction, for example $\varphi = 0$, the effective free energy is given by
\begin{equation}
    F_\varphi 
    \approx\int \mathrm{d}^2 r \left(
    \frac{\kappa}{2} (\boldsymbol{\nabla} \varphi)^2 + \frac{\mu_\perp m_0^2}{2} \varphi^2
    \right),
\end{equation}
in which $\mu_\perp > 0$ is the effective transverse pinning strength from the sixfold anisotropy.
The associated characteristic length is 
\begin{math}
    \xi_{\mu_\perp}^2 = {\kappa}/{(\mu_\perp m_0^2)}
    =
    K_m/\mu_\perp.
\end{math}
We obtain the estimate for the pinning strength,
\begin{equation}
    \mu_\perp
    \sim
    \frac{\kappa}{m_0^2 \xi_{\mu_\perp}^2}
    \sim
    \frac{K_m}{a^2}
    \exp
    \left(
    - \frac{2b}{\sqrt{(T_{\mu_\perp} - T)/T_{\mu_\perp}}}
    \right).
    \label{mu_perp.onset}
\end{equation}
The pinning is exponentially small near $T_{\mu_\perp}$, but nonzero for $T < T_{\mu_\perp}$.
Additional symmetry-breaking fields, such as an in-plane Zeeman field, also contribute to $\mu_\perp$.

The finite pinning cuts off the smectic regime by restoring a finite long-wavelength phase stiffness.
Namely, the averaged long-wavelength superfluid stiffness entering the Nelson-Kosterlitz scaling~\cite{Nelson1977} is given by~\cite{SM}
\begin{equation}
    \bar{\rho}_\mathrm{eff}(0)
    =
    \sqrt{\rho_\mathrm{eff}^x(0) \rho_\mathrm{eff}^y(0)}
    \approx
    \rho_s \sqrt{\frac{\mu_\perp}{\mu_\perp + \rho_s \lambda^2}},
    \label{effective_SF_stiffness.long_wavelength.with_pinning}
\end{equation}
for $\mu_\parallel \gg \rho_s \lambda^2$.
Consequently, for $r \gg \xi_{\mu_\perp}$, the transverse spin fluctuations are massive, and the vortex energy cost logarithmically diverges with system size, yielding a finite BKT transition temperature,
\begin{math}
    T_\mathrm{BKT} \lesssim (\pi/2)\bar{\rho}_\mathrm{eff}(0).
\end{math}
In other words, the pinning field cuts off the smectic-like behaviour at long wavelengths and restores a phase-coherent superconducting state at finite temperature.

\begin{figure}
    \centering
    \includegraphics[width=\linewidth]{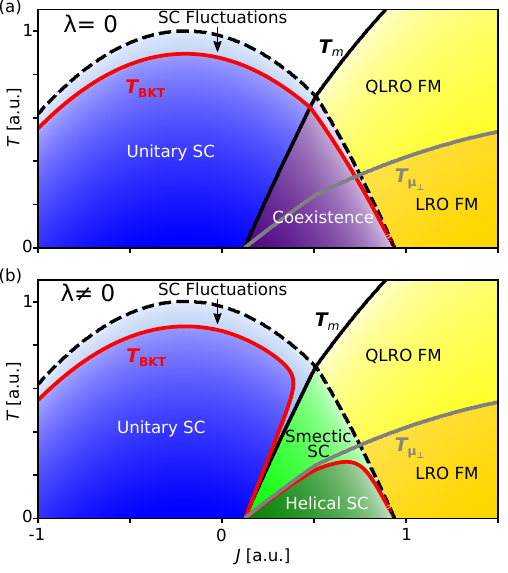}
    \caption{
    Phase diagrams for a Rashba superconductor coexisting with easy-plane ferromagnetism for magnetoelectric coupling (a)~$\lambda=0$ and (b)~$\lambda \neq 0$, as a function of temperature and phenomenological parameter, $J$. The ferromagnet is assumed to have an in-plane sixfold anisotropy.  
    Dark blue denotes a uniform unitary superconductor, and
    the light and dark yellow phases denote the 
QLRO
    and LRO 
canted ferromagnetic (FM) phases, respectively.
The solid red line denotes the BKT transition, and above, light shading indicates a vortex-deconfined regime. 
    The solid black contour denotes $T_m$ which describes the onset of QLRO FM, and the solid gray line, $T_{\mu_\perp}$, separates the QLRO FM from the LRO FM.
For $\lambda = 0$, the coexistence region shown in purple (panel a) is a nonunitary spin-polarized superconductor with zero center-of-mass momentum pairing.
    For $\lambda \neq 0$ (panel b), the same coexistence region realizes a smectic superconductor (light green) with zero long-wavelength phase stiffness, or below $T_{\mu_\perp}$, a QLRO helical nonunitary superconductor (dark green).
Dashed lines denote crossovers. 
    Details on the derivation of the phase diagram are included in the end matter.
    }
    \label{fig:phase_diagram}
\end{figure}

 \phantomsection
\addcontentsline{toc}{section}{Phase diagram of a smectic superconductor}
\paragraph*{{Phase diagram of a smectic superconductor.---}}We next discuss how the magnetoelectric coupling reshapes the finite-temperature phase diagram.
We consider a qualitative phase diagram as a function of temperature and a phenomenological parameter $J$, which controls the onset of canted in-plane ferromagnetism. 
The phase diagram contains a superconducting dome that partially overlaps with a canted ferromagnetic phase, as shown in Fig.~\ref{fig:phase_diagram}.
The relative placement of the superconducting and magnetic domes is motivated by rhombohedral graphene, where experiments suggest a possible regime in which superconductivity and canted ferromagnetism can coexist~\cite{Zhang2026}.

We first consider the case of vanishing magnetoelectric coupling, $\lambda = 0$.
In this limit, the phase diagram reduces to three distinct regions: a uniform unitary superconductor, a canted ferromagnet, and a region of coexistence, as shown in Fig.~\ref{fig:phase_diagram}(a).
Prior to the onset of finite spin polarization, the superconducting pairing order is unitary and spatially uniform.
This may be viewed as a superconducting state favored by spin-valley locking from, for example, Ising spin-orbit coupling.
Inside the coexistence region, the canted ferromagnetic order polarizes the Cooper pairs and thereby favors nonunitary pairing. 
In the absence of Rashba spin-orbit coupling, the nonunitary superconducting state remains spatially uniform, with zero center-of-mass momentum pairing.
Above $T_{\mu_\perp}$, the pinning $\mu_\perp$ vanishes, and the in-plane ferromagnetic order has QLRO, as shown in light yellow in Fig.~\ref{fig:phase_diagram}.
At $T_{\mu_\perp}$, the system undergoes a BKT-like transition, and the onset of finite pinning $\mu_\perp$ from lattice anisotropy selects a discrete easy-axis.
Consequently, for $T < T_{\mu_\perp}$, there is long-ranged-ordered easy-axis ferromagnetic order, as shown in dark yellow.
Because the soft magnetic fluctuations do not affect the superconducting phase stiffness for $\lambda=0$ , the BKT transition is not suppressed, and $T_\mathrm{BKT}$ roughly follows the onset of finite pairing correlations.

For finite magnetoelectric coupling ($\lambda \neq 0$), the coexistence region is qualitatively different, as shown in Fig.~\ref{fig:phase_diagram}(b).
The spin polarization couples linearly to the $\mathrm{U}(1)$ phase gradients through the Lifshitz invariant in Eq.~\eqref{ME_coupling}.
Once ferromagnetic order condenses, this coupling drives a helical superconducting state with finite center-of-mass momentum, $\mathbf{q}_0 = \lambda \hat{\mathbf{z}} \times \mathbf{m}_0$.
Above the pinning scale $T_{\mu_\perp}$, the in-plane magnetization retains an approximately rotationally symmetric character.
Soft transverse fluctuations of $\mathbf{m}(\mathbf{r})$ screen one component of the long-wavelength superconducting phase stiffness, 
driving a smectic superconducting regime, shown in light green in Fig.~\ref{fig:phase_diagram}(b).
In this regime, the vortex cost is no longer logarithmically divergent, and the BKT transition is suppressed at the onset of ferromagnetic QLRO.
In the absence of pinning, the finite-$q$ superconducting state retains finite local pairing amplitude but fails to have finite-temperature QLRO.
Below $T_{\mu_\perp}$, the sixfold anisotropy of the magnetic order generates a transverse pinning mass which cuts off the smectic softness at long wavelengths and restores a finite anisotropic phase stiffness.
The system can then undergo a finite-temperature BKT transition into a helical nonunitary superconductor with QLRO, shown in dark green in Fig.~\ref{fig:phase_diagram}(b).
This phase is distinguished from the $\lambda = 0$ coexistence phase in Fig.~\ref{fig:phase_diagram}(a) by its finite center-of-mass pairing momentum and anisotropic phase stiffness.

A vortex-deconfined region appears naturally near the onset of ferromagnetic order within the superconducting phase, \textit{i.e.} near the boundary between unitary and nonunitary, helical superconducting states.
There, the transverse magnetic susceptibility diverges, and the long-wavelength stiffness in Eq.~\eqref{effective_stiffness} falls below the BKT threshold.
This stiffness collapse is analogous to the vortex deconfinement near Ising criticality in a semidirect Ising-$XY$ model discussed in Ref.~\citenum{Yi-Thomas2026}.
There, vortices deconfine through their coupling to proliferating Ising domain walls, whereas in the present work, 
the soft mode is not a discrete Ising variable but rather a continuous easy-plane angle of the in-plane ferromagnetic order.
Consequently, the unpinned coexistence phase exhibits anisotropic, smectic-like phase elasticity rather than the typical vortex-deconfined response.
A finite pinning scale $\mu_\perp$
cuts off the divergence at long wavelengths and can stabilize QLRO in the superconducting phase.
Nonetheless, for any finite $\lambda$, a deconfined strip will remain, separating the unitary and nonunitary superconducting regimes, although its width and associated crossover scales can become parametrically small.

The smectic regime is characterized by two natural length scales,
\begin{math}
    \xi_\lambda = \sqrt{K_m/{\rho_s \lambda^2}}
\end{math}
and
\begin{math}
    \xi_{\mu_\perp} = \sqrt{K_m/{\mu_\perp}}.
\end{math}
For weak pinning $\mu_\perp \ll \rho_s \lambda^2$, these scales are well separated.
The former scale $\xi_\lambda$ marks the onset of smectic behaviour, whereas the latter scale marks the cutoff of smectic behaviour.
At distances $ r \ll \xi_\lambda$, the magnetic stiffness dominates over the magnetoelectric scale, and the system behaves as an ordinary $XY$ superfluid with stiffness $\rho_s$.
At intermediate distances $\xi_\lambda \ll r \ll \xi_{\mu_\perp}$, the transverse fluctuations screen phase gradients along one direction, leading to smectic-like elasticity.
Finally, at distances $r \gg \xi_{\mu_\perp}$, the pinning term cuts off the smectic behaviour, and the system has the behaviour of an anisotropic $XY$ model, with logarithmically confined vortices.

 \phantomsection
\addcontentsline{toc}{section}{Discussion}
\paragraph*{{Discussion.---}} 

We have shown that soft nonunitary fluctuations can couple directly to the superconducting $\mathrm{U}(1)$ phase, strongly reshaping phase coherence in a two-dimensional spin-orbit-coupled superconductor.
This coupling corresponds to a Lifshitz invariant permitted by broken inversion symmetry, with the in-plane spin polarization acting as an emergent gauge field for the $\mathrm{U}(1)$ phase.
In the coexistence regime, this gives rise to a helical $\mathrm{U}(1)$ texture, and
transverse nonunitary fluctuations screen the long-wavelength stiffness, driving the system into a smectic-like regime.
As a result, vortices lose their logarithmic cost, and near the onset of ferromagnetic order, there is a vortex-deconfined regime.
Weak in-plane pinning from lattice anisotropy cuts off the smectic-like elasticity at long distances, restoring finite BKT transition temperature.

We note finite size systems can cut off the divergent magnetic susceptibility, thus rounding the collapse of the superfluid stiffness.
To leading order, the magnetoelectric coupling gives a correction to the superfluid stiffness, $\delta \rho_s(L) \sim -(\rho_s \lambda)^2 \chi(L)$, while the magnetic QLRO gives susceptibility $\chi(L) \sim L^{2 - \eta(T)}$.
At the upper magnetic BKT transition, $\eta = 1/4$, giving~\cite{Kosterlitz1974, Hasenbusch2005} $\chi(L) \sim L^{7/4} (\ln (L/a) + C)^{1/8}$.
At the lower sixfold locking transition, $\eta = 1/9$, and the leading power is~\cite{Jose1977} $\chi(L) \sim L^{17/9}$.
Consequently, finite samples retain an apparent finite superfluid stiffness for $L \ll L_\lambda$, in which the crossover scale $L_\lambda$ is defined by $\rho_s \lambda^2 \chi(L_\lambda) \sim 1$.

The mechanism in this work is particularly relevant to rhombohedral graphene on a TMD substrate, where proximity-induced Rashba spin-orbit coupling and nearby canted ferromagnetic order provide the necessary ingredients for the magnetoelectric coupling.
In this setting, soft magnetic fluctuations can generate a smectic-like superconducting regime with vanishing long-wavelength stiffness, while an in-plane exchange field or intrinsic anisotropy that pins the ferromagnetic order can restore superconducting QLRO in the coexistence region.
More broadly, the results of this work demonstrate how soft fluctuations of the internal pairing structure can suppress phase coherence and qualitatively modify vortex energetics.

\vspace{2em}
\phantomsection
\addcontentsline{toc}{section}{Acknowledgements}
\paragraph*{{Acknowledgements.---}}
We thank Jay D. Sau for useful discussions on vortex-deconfined phases in a related semidirect Ising-$XY$ model, Leo Radzihovsky for insight into the smectic-like elasticity, and Andrea Young for helpful discussions on spin-orbit coupling in rhombohedral graphene proximitized by TMDs.

G.R.F. acknowledges partial support from NSF CAREER Grant No.~DMR-1848349, as well as the hospitality and support of the Kavli Institute for Theoretical Physics,
supported in part by the Heising-Simons Foundation, the Simons Foundation, and grant No. NSF PHY-2309135.
E.B. acknowledges funding support from the Simons Foundation Collaboration on New Frontiers in Superconductivity (Grant SFI-MPS-NFS-00006741-03), NSF-BSF Award No. DMR-2310312, and CRC 183 of the Deutsche Forschungsgemeinschaft (Project C02). 
This work was supported by a research grant from Magnus Konow in honour of his mother Olga Konow Rappaport.

\onecolumngrid
\phantomsection
\addcontentsline{toc}{section}{End Matter}
\begin{center}
    \textbf{End Matter}
\end{center}
\twocolumngrid

\section*{Phenomenological Phase Diagram}
\label{end_matter:phase_diagram}

In this section, we provide details about the minimal phenomenological model used in constructing the phase diagram of the smectic superconductor in Fig.~\ref{fig:phase_diagram}.
The model accounts for the interplay between superconducting order, in-plane ferromagnetism, magnetoelectric coupling permitted by broken inversion symmetry, and weak pinning of the in-plane magnetic order.

We write the superconducting order parameter minimally as
\begin{math}
    \Delta(\mathbf{r}) = |\Delta| e^{i \theta(\mathbf{r})},
\end{math}
and denote the in-plane magnetization by
\begin{math}
    \mathbf{m} = (m_x, m_y)^\mathrm{T} = m_0 \hat{\mathbf{x}} + \delta \mathbf{m},
\end{math}
in which $\delta \mathbf{m}$
denotes both transverse and longitudinal fluctuations.
The gauge-invariant phase gradient is given by
\begin{math}
    \mathbf{q} = \boldsymbol{\nabla} \theta - \frac{2e}{\hbar} \mathbf{A}_\mathrm{EM}.
\end{math}
Here, for simplicity, we consider the pairing amplitude and $\mathrm{U}(1)$ phase, and suppose that the $\mathrm{U}(1)$ phase reorients to minimize the free energy.
The field $\mathbf{m}$ phenomenologically represents the in-plane spin polarization inherited from the spin-canted normal state.

The free energy density can be decomposed as
\begin{equation}
    f = f_\Delta + f_m + f_\mathrm{mix} + f_{\mu_\perp}.
\end{equation}
The superconducting contribution is given by
\begin{equation}
    f_\Delta = \frac{r_\Delta}{2} |\Delta|^2
    + \frac{u_\Delta}{4} |\Delta|^4 + \frac{\rho_s}{2} |\mathbf{q}|^2,
\end{equation}
in which the bare superfluid stiffness is given by $\rho_s = \rho_0 |\Delta|^2$.
The magnetic contribution is
\begin{equation}
    f_m = \frac{r_m}{2} |\mathbf{m}|^2 + \frac{u_m}{4} |\mathbf{m}|^4,
\end{equation}
and the mixing is described by
\begin{equation}
    f_\mathrm{mix} = \frac{w}{2} |\Delta|^2 |\mathbf{m}|^2 + \lambda \rho_s \hat{\mathbf{z}} \cdot (\mathbf{q} \times \mathbf{m}) + \frac{\rho_s \lambda^2}{2} |\mathbf{m}|^2.
\end{equation}
The first term above describes the direct competition between the superconducting and magnetic amplitudes, with $w^2 < u_\Delta u_m$ ensuring that the quartic theory permits a stable coexistence in which both $|\Delta|$ and $m$ condense~\cite{Khomskii2010}.
The second and third terms emerge from the Lifshitz-type magnetoelectric coupling in Eq.~\eqref{ME_coupling}.
Finally, the magnetic pinning is given by
\begin{equation}
    f_\mathrm{\mu_\perp}
    =
    \frac{\mu_\perp^{(0)}}{2} m_y^2,
\end{equation}
and for $\mu_\perp^{(0)} > 0$, this term selects the $x$ direction as the easy-axis, assumed to be selected out of the six possible directions of the magnetic moment.

\phantomsection
\addcontentsline{toc}{subsection}{Saddle points and coexistence region}
\paragraph*{Saddle points and coexistence region.---}
The quadratic coefficients in the free energy are chosen phenomenologically to produce overlapping superconducting and magnetic domes, motivated by recent results in rhombohedral graphene~\cite{Zhang2026}.
We parameterize them as
\begin{subequations}
    \begin{align}
        r_\Delta(T, J)
        &=
        T - [a_\Delta - b_\Delta(J-c_\Delta)^2],
        \\
        r_m(T, J)
        &=
        T - \mathrm{max}[a_m - b_m(J- c_m)^2, 0].
    \end{align}
\end{subequations}
Here, $J$ is a phenomenological parameter describing the tendency towards superconducting and magnetic order.
The parameters $a_\Delta$ and $a_m$ set the maximal bare transition temperature, $b_\Delta$ and $b_m$ determine the curvatures, and $c_\Delta$ and $c_m$ determine the centers of the corresponding dome.
For the phase diagram in Fig.~\ref{fig:phase_diagram}, we use $a_\Delta = 1$,
$b_\Delta = 0.60$,
$c_\Delta = -0.20$,
$a_m = 1.7$,
$b_m = 0.35$,
and $c_m = 2.20$.
The remaining parameters are chosen as $u_\Delta = u_m = 1$, $w = 0.20$, $\rho_0 = 5.5$, and $\lambda = 0.40$.
The qualitative aspects of the phase diagram are relatively insensitive to the specific values of these parameters, provided there is a region of coexistence, with $w^2 < u_m u_\Delta$.

At the saddle point,
finite pinning selects $\mathbf{m} = m_0 \hat{\mathbf{x}}$, and the phase-dependent contribution is minimized by
\begin{equation}
    \mathbf{q}_0 = \lambda \hat{\mathbf{z}} \times \mathbf{m}_0 = \lambda m_0 \hat{\mathbf{y}},
\end{equation}
such that the gradient terms vanish.
For finite pinning, this direction is selected explicitly, whereas in the unpinned limit it may be chosen without loss of generality.
The relaxed homogeneous free energy is given by
\begin{equation}
    f_\mathrm{homog}
    =
    \frac{r_\Delta}{2} |\Delta|^2 + \frac{u_\Delta}{4} |\Delta|^4
    +
    \frac{r_m}{2} m_0^2 + \frac{u_m}{4} m_0^4
    +
    \frac{w}{2} |\Delta|^2 m_0^2.
\end{equation}
As such, the direct coupling $w$ determines the competition and coexistence of the two amplitudes.
The magnetoelectric coupling $\lambda$ instead determines the equilibrium superconducting momentum and fluctuation response, and $\mu_\perp^{(0)}$ gaps the transverse magnetic fluctuations.
At each point, the saddle point selects one of three possible nontrivial orders:
(i)~purely ferromagnetic order, with $m_0^2 = -r_m/u_m$,
(ii)~the unitary superconducting condensate with $|\Delta|^2 = -r_\Delta/u_\Delta$,
or
(iii) the coexisting nonunitary phase, with
$|\Delta|^2 = (w r_m - u_m r_\Delta)/(u_\Delta u_m - w^2)$
and
$m_0^2 = (w r_\Delta - r_m u_\Delta)/(u_\Delta u_m - w^2)$.
For $\lambda \neq 0$, the coexistence state has finite equilibrium phase gradient, with $q_{0, y} = \lambda m_0$.

We model the pinning of the magnetic fluctuations phenomenologically by allowing the bare transverse mass $\mu_\perp^{(0)}$ to become finite below a lower BKT-like transition nested within the magnetically ordered phase.
For each $J$, let $T_m(J)$ denote the onset of magnetic QLRO.
For simplicity, we model the pinning as
$\mu_\perp^{(0)}(T, J) = g_{\mu_\perp} \mathrm{max}[T_{\mu_\perp}(J)-T, 0]$, where we assume $T_{\mu_\perp} = \alpha T_m$, with $\alpha \leq 1$.
While this captures the qualitative behaviour, a more accurate description would include modelling the exponential behaviour in Eq.~\eqref{mu_perp.onset}.
In Fig.~\ref{fig:phase_diagram}, we take $\alpha = 0.35$ and $g_{\mu_\perp} = 0.12$.
This phenomenological transverse mass represents the easy-axis pinning of the magnetic order below the transition $T_{\mu_\perp}$ discussed in the main text.
The pinning does not modify the amplitudes of the pairing order but rather controls the energy cost of transverse fluctuations in $\mathbf{m}$.
Above $T_{\mu_\perp}$, the magnetic orientation remains unpinned and exhibits QLRO, whereas below $T_{\mu_\perp}$, transverse fluctuations are gapped and the remaining discrete easy-axis order can develop true LRO.

\begin{figure}[tb]
    \centering
    \includegraphics[width=\linewidth]{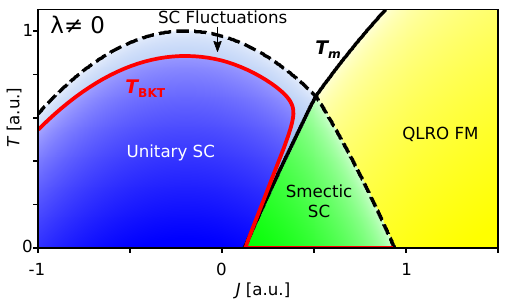}
    \caption{Phase diagram in the absence of pinning of the in-plane ferromagnetic order.
    Because the transverse magnetic mode remains gapless, the estimated BKT transition is suppressed to zero temperature.
    For $\lambda \neq 0$, the smectic superconducting phase occupies the entire finite-temperature region of coexistence.
    Apart from the absence of pinning ($T_{\mu_\perp^{(0)}} = 0$), the colors, labels, and parameters are same as those in Fig.~\ref{fig:phase_diagram}(b).}
    \label{fig:phase_diagram.no_pinning}
\end{figure}

\phantomsection
\addcontentsline{toc}{subsection}{Effective superfluid stiffness and BKT criterion}
\paragraph*{Effective superfluid stiffness and BKT criterion.---}
Upon expanding the magnetization about the homogeneous saddle points as $m_x = m_0 + \delta m_x$ and $m_y = \delta m_y$, it follows that the corresponding uniform magnetic curvatures are given by
\begin{subequations}
    \begin{align}
        \mu_\parallel 
        &= 
        r_m + 3 u_m m_0^2 + w |\Delta|^2,
        \\
        \mu_\perp 
        &= 
        r_m + u_m m_0^2 + w |\Delta|^2 + \mu_\perp^{(0)}.
    \end{align}
\end{subequations}
Within the magnetically ordered phase, $r_m + u_m m_0^2 + w|\Delta|^2 = 0$, and the curvatures are given by
$\mu_\parallel \approx 2 u_m m_0^2$ and $\mu_\perp \approx \mu_\perp^{(0)}$ to lowest order in $\delta \mathbf{m}$.

Expanding the phase gradient about its equilibrium value 
$\mathbf{q} = \lambda m_0 \hat{\mathbf{y}} + \delta \mathbf{q}$ and minimizing the quadratic free energy over $\delta \mathbf{m}$,
the resulting phase-only effective free energy is given by
\begin{equation}
    f^{(2)}_\mathrm{eff} 
=
    \frac{1}{2} \rho_\mathrm{eff}^x \delta q_x^2 + \frac{1}{2} \rho_\mathrm{eff}^y \delta q_y^2,
\end{equation}
in which
\begin{equation}
    \rho_\mathrm{eff}^x
    =
    \frac{\rho_s \mu_\perp}{\mu_\perp + \rho_s \lambda^2},
    \hspace{2em}
    \rho_\mathrm{eff}^y
    =
    \frac{\rho_s \mu_\parallel}{\mu_\parallel + \rho_s \lambda^2}.
\end{equation}
The effective stiffness entering the BKT criterion~\cite{Nelson1977} is given by
\begin{equation}
    \bar{\rho}_\mathrm{eff}
    =
    \sqrt{\rho_\mathrm{eff}^x \rho_\mathrm{eff}^y}
    =
    \rho_s
    \sqrt{
    \frac{\mu_\perp \mu_\parallel}{(\mu_\parallel + \rho_s \lambda^2)(\mu_\perp + \rho_s \lambda^2)}}.
\end{equation}
We estimate the BKT transition from the condition $\bar{\rho}_\mathrm{eff}(T_\mathrm{BKT}) = 2 T_{\mathrm{BKT}}/\pi$.
Below $T_\mathrm{BKT}$, vortices are bound and the superconducting state has QLRO, whereas above $T_\mathrm{BKT}$, vortices deconfine.
For $\lambda \neq 0$, and an unpinned easy-plane ferromagnet with QLRO ($\mu_{\perp}^{(0)} = 0$), the effective long-wavelength stiffness vanishes, corresponding to the smectic superconducting regime discussed in the main text.

In Fig.~\ref{fig:phase_diagram.no_pinning}, we show the phase diagram for $T_{\mu_\perp} = 0$, corresponding to the unpinned easy-plane ferromagnetic order at finite temperature.
In contrast to Fig.~\ref{fig:phase_diagram}(b), in which $T_{\mu_\perp} \neq 0$ and there is finite pinning which allows for a vortex-confined helical pairing phase, the finite-temperature coexistence region is fully occupied by the vortex-deconfined smectic superconducting phase (light green).

Lastly, a general free energy also contains gradient terms in magnetic order, as introduced in Eq.~\eqref{free_energy_density}.
This term does not affect the homogeneous saddle points and as such is not included in this simple phenomenological model.
It is, however, essential for the finite-wave-vector response in the limit of weak or vanishing pinning,
as shown in Eq.~\eqref{effective_stiffness}.
This gives the smectic-like elasticity and vortex cost discussed in the main text.

\end{document}


\title{
Supplemental Material for ``Smectic Superconductivity''
}

\author{Grayson R. Frazier}
\affiliation{Department of Physics and Astronomy, Johns Hopkins University, Baltimore, Maryland 21218, USA}
\affiliation{Kavli Institute for Theoretical Physics, University of California, Santa Barbara, CA 93106, USA}
\author{Erez Berg}
\affiliation{Department of Condensed Matter Physics, Weizmann Institute of Science, Rehovot 76100, Israel}
\affiliation{Materials Department, University of California Santa Barbara, Santa Barbara 93106 USA}
\affiliation{Department of Electrical and Computer Engineering, University of California, Santa Barbara, CA 93106, USA}

\onecolumngrid

\setcounter{equation}{0}
\setcounter{section}{0}
\setcounter{figure}{0}
\setcounter{table}{0}
\setcounter{page}{1}
\makeatletter
\renewcommand{\theequation}{S\arabic{equation}}
\renewcommand{\thefigure}{S\arabic{figure}}
\renewcommand{\thesection}{S\arabic{section}}
\renewcommand{\thepage}{S\arabic{page}}

\onecolumngrid

\newpage

\clearpage
\setcounter{secnumdepth}{2}

\setcounter{equation}{0}
\setcounter{section}{0}
\setcounter{figure}{0}
\setcounter{table}{0}
\setcounter{page}{1}
\makeatletter
\renewcommand{\theequation}{S\arabic{equation}}
\renewcommand{\thefigure}{S\arabic{figure}}
\renewcommand{\thesection}{S\arabic{section}}

\phantomsection
\addcontentsline{toc}{title}{Supplemental Material}
\begin{center}
    \textbf{\large{Supplemental Material for
    ``Smectic Superconductivity''
    }}
    \\
    \vspace{1em}
    Grayson R. Frazier$^{1,2}$ and Erez Berg$^{3,4,5}$
    \\
    \textit{\small $^1$Department of Physics and Astronomy, Johns Hopkins University, Baltimore, Maryland 21218, USA}
    \\
    \textit{\small $^2$Kavli Institute for Theoretical Physics, University of California, Santa Barbara, CA 93106, USA}
    \\
    \textit{\small $^3$Department of Condensed Matter Physics, Weizmann Institute of Science, Rehovot 76100, Israel
    }
    \\
    \textit{\small $^4$Materials Department, University of California Santa Barbara, Santa Barbara 93106 USA}
    \\
    \textit{\small $^5$Department of Electrical and Computer Engineering, University of California, Santa Barbara, CA 93106, USA}

\end{center}

 \section{Magnetoelectric Coupling and Smectic-Like Elasticity}

We consider superconducting pairing order given by
\begin{equation}
    \hat{\Delta}_{\alpha \beta}(\mathbf{r}, \mathbf{k})
    =
    e^{i\theta(\mathbf{r})}
    [(
    \psi(\mathbf{r}, \mathbf{k})
    +
    \mathbf{d}(\mathbf{r}, \mathbf{k}) \cdot \boldsymbol{\sigma}
    )i  \sigma^y]_{\alpha \beta}.
\end{equation}
Here, $\alpha, \beta  = \uparrow, \downarrow$ are spin indices, $\theta(\mathbf{r})$ is the $\mathrm{U}(1)$ superconducting phase, $\sigma^i$ are Pauli spin matrices, $\mathbf{r}$ is the center-of-mass coordinate of the Cooper pair, and $\mathbf{k}$ is the relative momentum.
The field $\psi(\mathbf{r}, \mathbf{k}) = \psi(\mathbf{r}, -\mathbf{k})$ describes spin singlet pairing correlations, and $\mathbf{d}(\mathbf{r}, \mathbf{k}) = -\mathbf{d}(\mathbf{r}, -\mathbf{k})$ is the $d$ vector describing spin triplet pairing correlations.
In the following, we discuss the possible magnetoelectric coupling arising from nonunitary fluctuations in the $d$ vector field and show how this can lead to a smectic-like elasticity.

\subsection{Nonunitary Pairing Fluctuations and Magnetoelectric Coupling}

We study the minimal low-energy theory of a superconductor proximate to easy-plane, spin-polarized nonunitary pairing order.
Near the onset of coexisting ferromagnetic order, fluctuations which generate the spin polarization of the superconducting condensate become soft and can therefore strongly affect the effective superfluid stiffness.

For simplicity, we consider the spin triplet channel to be described by
\begin{math}
    \mathbf{d}(\mathbf{r}, \mathbf{k}) 
    =
    g(\mathbf{k}) \hat{\mathbf{d}}(\mathbf{r}),
\end{math}
where any fixed overall amplitude is absorbed into the orbital factor $g(\mathbf{k})$, and we take 
$|\hat{\mathbf{d}}| = 1$.
To describe fluctuations of the triplet pairing state, we hold the gap amplitude and orbital pairing structure fixed, and we parameterize the normalized $d$ vector as
\begin{align}
    \hat{\mathbf{d}}(\mathbf{r})
    &= \nonumber
    \sqrt{1 - \beta^a(\mathbf{r})
    \beta^a(\mathbf{r})
    }
    \hat{\mathbf{n}}_0 (\mathbf{r})
    +
    i \beta^a(\mathbf{r}) \hat{\mathbf{t}}_a (\mathbf{r})
    \\
    &\approx
    \left(1 - \frac{1}{2} \beta^a(\mathbf{r}) \beta^a(\mathbf{r}) \right)
    \hat{\mathbf{n}}_0  (\mathbf{r})
    +
    i \beta^a(\mathbf{r})
    \hat{\mathbf{t}}_a (\mathbf{r}),
\end{align}
in which summation is implied over repeated indices $a = 1, 2$.
Here, 
$\hat{\mathbf{n}}_0(\mathbf{r})$ 
is the axis of the unitary $d$ vector, which we assume to be pinned by, for example, Ising spin-orbit coupling.
$\beta^a(\mathbf{r}) \in \mathbb{R}$ corresponds to soft nonunitary fluctuations, with $|\beta^a(\mathbf{r})| \ll 1$, and $\{\hat{\mathbf{n}}_0 (\mathbf{r}), \hat{\mathbf{t}}_1 (\mathbf{r}), \hat{\mathbf{t}}_2 (\mathbf{r})\}$ forms a local orthonormal triad.
Near the onset of spin-polarized nonunitary superconducting order, the low-energy fluctuations are imaginary components of $\hat{\mathbf{d}}$ transverse to $\hat{\mathbf{n}}_0$, which generate the in-plane spin polarization,
\begin{equation}
    \mathbf{m} (\mathbf{r}) 
    \propto 
    i \hat{\mathbf{d}} (\mathbf{r}) \times \hat{\mathbf{d}}^* (\mathbf{r}) 
    = 
    2 \beta^a(\mathbf{r}) 
    \Big(\hat{\mathbf{n}}_0(\mathbf{r}) \times \hat{\mathbf{t}}_a (\mathbf{r})
    \Big)
    + \mathcal{O}((\beta^a)^3).
    \label{SM.Eq:magnetization}
\end{equation}
It is useful to define the transverse fluctuation field,
\begin{equation}
    \boldsymbol{\beta}(\mathbf{r}) \equiv \beta^a(\mathbf{r}) \hat{\mathbf{t}}_a(\mathbf{r}),
\end{equation}
with $\boldsymbol{\beta} \perp \hat{\mathbf{n}}_0$, which, to leading order, can be expressed as
\begin{equation}
    \boldsymbol{\beta} (\mathbf{r}) \propto - \hat{\mathbf{n}}_0 (\mathbf{r}) \times \mathbf{m} (\mathbf{r}).
\end{equation}

In a noncentrosymmetric superconductor, a bilinear coupling between the gauge-invariant phase gradient and spin polarization is allowed.
Here, the magnetoelectric coupling has the following form,
\begin{equation}
    f_\mathrm{ME}
    \propto
    \mathbf{q}(\mathbf{r}) \cdot \boldsymbol{\beta}(\mathbf{r})
    \propto
    \mathbf{q}(\mathbf{r}) \cdot \Big( \hat{\mathbf{n}}_0(\mathbf{r}) \times \mathbf{m}(\mathbf{r}) \Big).
\end{equation}
This term is even under time reversal, as both $\mathbf{q}$ and $\mathbf{m}$ are time-reversal odd.
For the system considered in the main text, the unitary component is oriented along $\hat{\mathbf{n}}_0 = \hat{\mathbf{z}}$, as in the case of Ising spin-orbit coupling.
Consequently, the nonunitary component and the magnetization in Eq.~\eqref{SM.Eq:magnetization} lie in the plane, $\mathbf{m} \perp \hat{\mathbf{z}}$.
In a two-dimensional noncentrosymmetric system with Rashba spin-orbit coupling, this gives rise to the leading magnetoelectric Lifshitz invariant,
\begin{equation}
    f_\mathrm{ME}
    \propto
    \mathbf{q}(\mathbf{r}) \cdot 
    \Big(
    \hat{\mathbf{z}} \times \mathbf{m} (\mathbf{r})
    \Big),
\end{equation}
as discussed in Eq.~\eqref{ME_coupling} of the main text.
While the above argument motivates the form of this coupling from symmetry, it can arise microscopically from Rashba spin-orbit coupling, as discussed in S.M. Sec.~\ref{SM:sec:R3G_ME_coupling}.

\subsection{Effective Free Energy and Smectic-like Elasticity}

We consider the following phenomenological free energy,
\begin{equation}
    F[\theta, \mathbf{m}]
    =
    \frac{1}{2}
    \int \mathrm{d}^2 r
    \Big[
    \rho_s
    (\mathbf{q} - \lambda \hat{\mathbf{z}} \times \mathbf{m})^2
    +
    K_m (\partial_i \mathbf{m}) \cdot (\partial_i \mathbf{m})
    +
    r_m \mathbf{m}^2
    +
    \frac{1}{2} u_m (\mathbf{m}^2)^2
    +
    \mu_\perp^{(0)} m_y^2
    -
    2\mathbf{h}_\parallel \cdot \mathbf{m}
    \Big],
    \label{free_energy_with_coupling}
\end{equation}
in which $\rho_s$ is the bare superfluid stiffness, $K_m$ is the magnetic stiffness, and $r_m$ and $u_m$ are phenomenological coefficients describing the onset of magnetic order.
Here, $\mathbf{q} (\mathbf{r}) = \boldsymbol{\nabla} \theta(\mathbf{r}) - ({2\pi}/{\Phi_0}) \mathbf{A}_\mathrm{EM}$ is the gauge-invariant $\mathrm{U}(1)$ phase gradient, with $\Phi_0 = h/2e$ the superconducting flux quantum.
The coefficient $\mu_\perp^{(0)}$ describes the explicit in-plane pinning of the magnetic order, which can arise from, for example, lattice anisotropy, as discussed in S.M.~Sec.~\ref{appendix:pinning}.
We also introduce an in-plane symmetry-breaking field $\mathbf{h}_\parallel$, such as an applied Zeeman field.
For simplicity,
we take $\mathbf{h}_\parallel = h_\parallel \hat{\mathbf{x}}$.

Expanding the first term yields the magnetoelectric Lifshitz invariant,
\begin{equation}
    f_\mathrm{ME}
    =
    \lambda \rho_s \hat{\mathbf{z}} \cdot (\mathbf{q} \times \mathbf{m}),
\end{equation}
which is permitted when inversion symmetry is broken---for example, in the presence of Rashba spin-orbit coupling~\cite{Edelstein1995, Agterberg2012}.
The in-plane magnetization therefore acts as an emergent gauge field for the superconducting phase.
Suppose that the magnetization condenses along the $x$ direction,
\begin{equation}
    \mathbf{m} = \mathbf{m}_0 + \delta \mathbf{m},
\end{equation}
with 
$ \mathbf{m}_0 = m_0 \hat{\mathbf{x}}$ and
$\delta \mathbf{m} = \delta m_x \hat{\mathbf{x}} + \delta m_y \hat{\mathbf{y}}$ denoting small fluctuations.
The equilibrium magnetization is determined by the saddle-point condition,
\begin{equation}
    r_m m_0 + u_m m_0^3 = h_\parallel.
    \label{saddle_point_condn}
\end{equation}
For the uniform, vortex-free saddle point, minimization with respect to $\mathbf{q}$ yields
\begin{equation}
    \mathbf{q}_0 = \lambda m_0 \hat{\mathbf{y}},
\end{equation}
and the corresponding equilibrium current vanishes, $\mathbf{j}_0 = \rho_s (\mathbf{q}_0 - \lambda \hat{\mathbf{z}} \times \mathbf{m}_0) = 0$.
We consider $\mathbf{q} = \mathbf{q}_0 + \delta \mathbf{q}$, where $\delta \mathbf{q}$ denotes fluctuations.
Near this saddle point, the effective free energy at the quadratic level is given by
\begin{equation}
    F^{(2)}
    =
    \frac{1}{2}
    \int \mathrm{d}^2 r
    \Big[
    \rho_s (\delta q_x + \lambda \delta m_y)^2
    +
    \rho_s (\delta q_y - \lambda \delta m_x)^2
    +
    K_m (\partial_i \delta \mathbf{m})^2
    +
    \mu_\parallel (\delta m_x)^2 + \mu_\perp (\delta m_y)^2
    \Big]
\end{equation}
in which longitudinal and transverse curvatures are given by
\begin{subequations}
    \begin{align}
        \mu_\parallel 
        &= r_m + 3 u_m m_0^2 
        = \frac{h_\parallel}{m_0} + 2 u_m m_0^2,
        \\
        \mu_\perp 
        &= r_m + u_m m_0^2 + \mu_\perp^{(0)}
        =
        \frac{h_\parallel}{m_0} + \mu_\perp^{(0)}.
    \end{align}
\end{subequations}
In the above equalities, we have used the saddle-point condition in Eq.~\eqref{saddle_point_condn}.
Consequently, $\delta m_x$ corresponds to a longitudinal fluctuation in $\mathbf{m}$ and is generically massive.
In contrast, $\delta m_y$ corresponds to a transverse rotation of the ordered moment.
In the absence of an explicit in-plane pinning ($\mu_\perp^{(0)} = h_\parallel = 0$), 
the transverse mode is the Goldstone mode associated with spontaneous breaking of the continuous in-plane rotation symmetry.

Taking the Fourier transform, the quadratic free energy is given by
\begin{equation}
    F^{(2)}
    =
    \frac{1}{2}
    \int \frac{\mathrm{d}^2 k}{(2\pi)^2}
    \Big[
    \rho_s |\delta \mathbf{q} (\mathbf{k})|^2 + M_\parallel(\mathbf{k})
    |\delta m_x(\mathbf{k})|^2 + M_\perp (\mathbf{k}) |\delta m_y(\mathbf{k})|^2 + 2\rho_s \lambda
    \Big(
    \delta q_x(-\mathbf{k}) \delta m_y(\mathbf{k}) - \delta q_y (-\mathbf{k}) \delta m_x (\mathbf{k})
    \Big)
    \Big],
\end{equation}
in which
\begin{subequations}
    \begin{align}
        M_\parallel(\mathbf{k}) 
        &= 
        K_m k^2 + \mu_\parallel + \rho_s \lambda^2,
        \\
        M_\perp(\mathbf{k})
        &=
        K_m k^2 + \mu_\perp + \rho_s \lambda^2.
    \end{align}
\end{subequations}
Integrating out Gaussian fluctuations $\delta \mathbf{m}$ yields the momentum-dependent effective phase stiffness,
\begin{subequations}
    \begin{align}
        \rho_\mathrm{eff}^x (\mathbf{k}) 
        &= 
        \rho_s - \frac{\rho_s^2 \lambda^2}{M_\perp(\mathbf{k})}
        =
        \rho_s \frac{K_m k^2 + \mu_\perp}{K_m k^2 + \mu_\perp + \rho_s \lambda^2},
        \\
        \rho_\mathrm{eff}^y (\mathbf{k}) 
        &= \rho_s - \frac{\rho_s^2 \lambda^2}{M_\parallel (\mathbf{k})}
        =
        \rho_s \frac{K_m k^2 + \mu_\parallel}{K_m k^2 + \mu_\parallel + \rho_s \lambda^2}.
    \end{align}
    \label{SM.Eq:effective_stiffness}
\end{subequations}
The resulting phase-only effective free energy is given by
\begin{align}
    F_\mathrm{eff}
    &=
    \frac{1}{2}
    \int \frac{\mathrm{d}^2 k}{(2\pi)^2}
    \Big[
    \rho_\mathrm{eff}^x (\mathbf{k}) |\delta q_x(\mathbf{k})|^2 + \rho_\mathrm{eff}^y(\mathbf{k}) |\delta q_y (\mathbf{k})|^2
    \Big].
\end{align}

When $\mu_\parallel \gg \rho_s \lambda^2$, the long-wavelength stiffness associated with phase gradients along the $y$ direction remains approximately equal to the bare stiffness, $\rho_\mathrm{eff}^y \approx \rho_s$.
In contrast, when there is full rotational symmetry in the magnetic order corresponding to easy-plane ferromagnetism ($\mu_\perp \rightarrow 0$), the gapless transverse magnetic fluctuations strongly renormalize the long-wavelength effective phase stiffness $\rho_\mathrm{eff}^x$ in Eq.~\eqref{SM.Eq:effective_stiffness}.
At long wavelengths $K_m k^2 \ll \rho_s \lambda^2$, this becomes
\begin{equation}
    \rho_\mathrm{eff}^x (\mathbf{k}) \approx \frac{K_m}{\lambda^2} k^2,
\end{equation}
leading to a quartic gradient term in the effective free energy.

For smooth phase fluctuations $\delta q_x = ik_x \delta \theta$ and $\delta q_y = ik_y \delta \theta$, the effective free energy is given by
\begin{equation}
    F_\mathrm{eff} \approx \frac{1}{2} \int \frac{\mathrm{d}^2 k}{(2\pi)^2}
    \left[
    \rho_s k_y^2 + \frac{K_m}{\lambda^2} k^2 k_x^2
    \right]
    |\delta \theta (\mathbf{k})|^2
    =
    \frac{1}{2} \int \mathrm{d}^2r
    \left[
    \rho_s (\partial_y \delta \theta)^2 + \frac{K_m}{\lambda^2} (\boldsymbol{\nabla} \partial_x \delta \theta)^2
    \right],
    \label{SMEq:smectic_analogy}
\end{equation}
which has the form of a two-dimensional smectic.
In the absence of explicit in-plane symmetry breaking, the helical saddle point spontaneously selects a direction, producing a spatially modulated phase texture.
The corresponding $x$-direction phase stiffness vanishes quadratically with momentum.
A finite transverse pinning $\mu_\perp > 0$, arising from either explicit in-plane anisotropy or a finite symmetry-breaking field, cuts off the smectic-like elasticity at sufficiently long wavelengths and restores a finite long-wavelength phase stiffness.

\subsection{Vortex Cost}

As shown in Eq.~\eqref{SMEq:smectic_analogy}, the system has an effective free energy resembling that of a smectic, which has been shown to lead to finite vortex cost and dislocation deconfinement~\cite{Toner1981, Gennes1993}.
For completeness,
we derive the energetic cost of a vortex defect.
For a vortex excitation described by
\begin{equation}
    \hat{\mathbf{z}} \cdot (\boldsymbol{\nabla} \times \boldsymbol{\nabla} \delta \theta
    )= 2\pi \delta(\mathbf{r}),
\end{equation}
we introduce an auxiliary field,
\begin{equation}
    \mathbf{e}(\mathbf{r})
    =
    \boldsymbol{\nabla} \delta \theta(\mathbf{r}) 
    \times
    \hat{\mathbf{z}}
    =
    \left(
    \begin{array}{c}
         \partial_y \delta \theta
         \\
         -\partial_x \delta \theta
    \end{array}
    \right).
\end{equation}
This is associated with the Gauss-law constraint,
\begin{equation}
    \boldsymbol{\nabla}
    \cdot \mathbf{e}
    =
    2\pi \delta(\mathbf{r}),
\end{equation}
or equivalently,
\begin{equation}
    i\mathbf{k} \cdot \mathbf{e}(\mathbf{k}) = 2\pi,
\end{equation}
in which $\mathbf{e}(\mathbf{k})$ is the Fourier-transformed field.

As such, the effective free energy can be rewritten as follows,
\begin{align}
    F_\mathrm{eff}
    &=
    \frac{1}{2}
    \int \mathrm{d}^2r
    \left(
    \frac{K_m}{ \lambda^2} (\boldsymbol{\nabla} \partial_x \delta \theta)^2
    +
    \rho_s
    (\partial_y \delta \theta)^2
    \right)
    =
    \frac{1}{2}
    \int \mathrm{d}^2r 
    \left(
    \frac{K_m}{ \lambda^2} (\boldsymbol{\nabla} e_y)^2
    +
    \rho_s
    (e_x)^2
    \right)
    =
    \frac{1}{2}
    \sum_\mathbf{k}
    \left(
    \frac{K_m}{\lambda^2}
    k^2 |e_y(\mathbf{k})|^2
    +
    \rho_s |e_x(\mathbf{k})|^2
    \right).
\end{align}
Minimizing with respect to the auxiliary field, the effective free energy is given by
\begin{equation}
    \tilde{F}
    =
    \sum_\mathbf{k}
    \left(
    \frac{K_m}{2\lambda^2} k^2
    |e_y(\mathbf{k})|^2 + \frac{\rho_s}{2}|e_x(\mathbf{k})|^2
    +
    \eta_\mathbf{k} 
    \Big(
    i k_x e_x(\mathbf{k}) + i k_y e_y(\mathbf{k}) - 2\pi
    \Big)
    \right).
\end{equation}
We define
\begin{equation}
    \mathbf{e}(\mathbf{k}) = i \boldsymbol{\varepsilon}(\mathbf{k}).
\end{equation}
Minimizing with respect to $\boldsymbol{\varepsilon}(\mathbf{k})$, it follows that
\begin{equation}
    \varepsilon_y(\mathbf{k}) =\frac{\lambda^2 k_y}{K_m k^2} \eta_\mathbf{k},
    \hspace{2em}
    \varepsilon_x(\mathbf{k}) =\frac{k_x}{\rho_s} \eta_\mathbf{k},
\end{equation}
and from the constraint,
\begin{equation}
    \eta_\mathbf{k} 
    = 
    - \frac{2\pi}{(\frac{k_x^2}{\rho_s}) + (\frac{\lambda^2 k_y^2}{K_m k^2})}.
\end{equation}
Substituting into the vortex cost,
it follows that
\begin{align}
    F_\mathrm{vortex}
    =
    \frac{1}{2}
    \sum_{\mathbf{k}} \left( \frac{K_m}{\lambda^2} k^2 \varepsilon_y^2 + \rho_s \varepsilon_x^2 \right)
    =
    \frac{1}{2}
    \sum_\mathbf{k}
    \frac{(2\pi)^2}{(\frac{k_x^2}{\rho_s})
    +
    (\frac{\lambda^2 k_y^2}{K_m k^2} )}.
\end{align}
Approximating the discrete sum as an integral,
\begin{align}
    F_\mathrm{vortex}
    =
    \frac{1}{2} (2\pi)^2
\int_{k_\mathrm{IR} < 
    k < \Lambda}
    \frac{\mathrm{d}^2 k}{(2\pi)^2}
    \frac{1}{(\frac{k_x^2}{\rho_s})
    +
    (\frac{\lambda^2 k_y^2}{K_m k^2} )}
    =
    \frac{1}{2}
\int_{k_\mathrm{IR} < k < \Lambda}
    k \, \mathrm{d}k
    \int\mathrm{d}\phi \frac{1}{(\frac{k^2 \cos^2\phi}{\rho_s})
    +
    (\frac{\lambda^2 \sin^2 \phi}{K_m})},
\end{align}
in which
$k_\mathrm{IR}$ and $\Lambda$ are IR and UV cutoffs, respectively.
Using the following integral,
\begin{equation}
    \int_0^{2\pi} \frac{\mathrm{d}\phi}{A\cos^2 \phi + B\sin^2 \phi}
    =
    \frac{2\pi}{\sqrt{AB}},
\end{equation}
it follows that
\begin{equation}
    F_\mathrm{vortex} = \frac{1}{2}
    \int_{k_\mathrm{IR}}^{\Lambda}
    \mathrm{d}k 
    \frac{2\pi \sqrt{K_m \rho_s}}{|\lambda|}
    =
    \frac{\pi\sqrt{K_m \rho_s}}{|\lambda|}
    (\Lambda - k_\mathrm{IR}).
\end{equation}
In the thermodynamic limit $L \rightarrow \infty$, we take $k_\mathrm{IR} \rightarrow 0$.
Thus, the vortex cost does not exhibit the usual logarithmic divergence but rather remains finite in the infrared.
This is analogous to the behaviour in a two-dimensional smectic, in which finite vortex cost leads to dislocation deconfinement.

\subsection{Emergent Length Scales}

\label{SM:subsec:length_scales}

In the presence of finite magnetoelectric coupling and weak pinning, 
there are two natural length scales.
The first is set by the magnetoelectric coupling $\lambda$ and can be seen as follows.
In the regime of full rotational symmetry of the magnetic order ($\mu_\perp = 0$), the effective stiffness in Eq.~\eqref{SM.Eq:effective_stiffness} for gradients in the $x$ direction is given by
\begin{equation}
    \rho_\mathrm{eff}^x (\mathbf{k}) = \rho_s \frac{K_m k^2}{K_m k^2 + \rho_s \lambda^2}.
\end{equation}
For short wavelengths $K_m k^2 \gg \rho_s \lambda^2$, the effective stiffness approaches the bare stiffness, 
\begin{equation}
\rho_\mathrm{eff}^x(\mathbf{k}) \approx \rho_s,
    \hspace{2em}
    \text{for $k \gg k _\lambda$}.
\end{equation}
In contrast, for long wavelengths $K_m k^2 \ll \rho_s \lambda^2$, and the effective stiffness is given by
\begin{equation}
   \rho_\mathrm{eff}^x(\mathbf{k}) \approx \frac{K_m k^2}{\lambda^2},
   \hspace{2em}
   \text{for $k \ll k_\lambda$},
\end{equation}
such that $\lim_{\mathbf{k} \rightarrow 0} \rho_\mathrm{eff}^x(\mathbf{k}) = 0$.
The ordinary $x$-direction phase stiffness vanishes in the infrared, and the system becomes smectic-like.
The crossover occurs when
\begin{equation}
    \rho_s \sim \frac{K_m k^2_\lambda}{\lambda^2},
\end{equation}
or equivalently, $k_\lambda^2 = \lambda^2 \rho_s/K_m$,
and
the associated length scale is
\begin{equation}
    \xi_\lambda \sim \frac{1}{k_\lambda}
    \sim\sqrt{\frac{K_m}{\rho_s \lambda^2}}.
\end{equation}
For distances shorter than $\xi_\lambda$, the system behaves as an ordinary $XY$ superfluid, but for longer distances, the smectic-like behaviour is important.

The second length scale is set by the pinning term $\mu_\perp$.
The magnetic fluctuations have quadratic energy
\begin{equation}
    \frac{1}{2} (K_m k^2 + \mu_\perp) |\delta m( \mathbf{k})|^2.
\end{equation}
The crossover between gradient-dominated and pinning-dominated behaviour occurs when
\begin{equation}
    K_m k^2_{\mu_\perp} \sim \mu_\perp,
\end{equation}
defining the second length scale
\begin{equation}
    \xi_{\mu_\perp} \sim \frac{1}{k_{\mu_\perp}} \sim \sqrt{\frac{K_m}{\mu_\perp}}.
\end{equation}
Beyond the scale $\xi_{\mu_\perp}$, pinning cuts off the smectic behaviour and restores the usual logarithmic vortex energy cost.

Thus, for weak pinning ($\mu_\perp \ll \rho_s \lambda^2$), it follows that $k_{\mu_\perp} \ll k_\lambda$, or equivalently, $\xi_{\mu_\perp} \gg \xi_\lambda$.
As such, the system can cross through three regimes:
(i)~$r \ll \xi_\lambda$: 
ordinary approximately isotropic $XY$ behaviour with stiffness $\rho_s$;
(ii)~$\xi_\lambda \ll r \ll \xi_{\mu_\perp}$:
smectic-like regime with finite vortex cost;
and
(iii)~$r \gg \xi_{\mu_\perp}$:
anisotropic $XY$ behaviour with logarithmic vortex cost.

 \section{Magnetoelectric Coupling in Rhombohedral Graphene}

\label{SM:sec:R3G_ME_coupling}

In this section, we derive the magnetoelectric coupling in rhombohedral graphene in proximity to a transition-metal dichalcogenide substrate.
We define the magnetoelectric coupling by expanding the free energy as
\begin{equation}
    F[\mathbf{q}, \mathbf{m}]
    =
    F[0,0] + \int \mathrm{d}^2 r
    \Big[
    \frac{1}{2}\rho_s^{ij} q_i(\mathbf{r}) q_j(\mathbf{r})
    +
    q_i(\mathbf{r}) \Lambda_{q_i,m_j} m_j(\mathbf{r})
    \Big]
    +
    \cdots,
\end{equation}
in which $\mathbf{m}(\mathbf{r})$ is a dimensionless vector describing the canted ferromagnetic order.
The uniform magnetoelectric response tensor is defined as~\cite{Abrikosov1963}
\begin{equation}
    \Lambda_{q_i, m_j}
    \equiv
    \left.\frac{\partial^2 f}{\partial q_i \partial m_j} \right|_{\mathbf{q}=0, \mathbf{m}=0},
\end{equation}
where $f$ is the free-energy density, defined by
$F[\mathbf{q}, \mathbf{m}] = \int \mathrm{d}^2r \, f[\mathbf{q}, \mathbf{m}]$.
For spatially varying fields,
the magnetoelectric contribution can be written as
\begin{equation}
    F_\mathrm{ME} =
    \sum_{\mathbf{k}} q_i(-\mathbf{k}) \Lambda_{q_i, m_j} (\mathbf{k})
    m_j(\mathbf{k}).
\end{equation}
We assume that the fields $\mathbf{q}(\mathbf{r})$ and $\mathbf{m}(\mathbf{r})$ are slowly varying on microscopic scales and evaluate the susceptibility for long wavelengths,
\begin{math}
\lim_{\mathbf{k} \rightarrow 0} \Lambda_{q_i, m_j}(\mathbf{k}).
\end{math}
In the main text, the magnetoelectric coupling is defined by
\begin{equation}
    f_\mathrm{ME} = \lambda \rho_s \hat{\mathbf{z}} \cdot (\mathbf{q} \times \mathbf{m}),
\end{equation}
corresponding to
\begin{equation}
    \lambda = \Lambda_{q_x, m_y}/\rho_s.
\end{equation}
In the following, we derive $\Lambda$ and $\rho_s$ from a continuum model of rhombohedral trilayer graphene.

\subsection{Model of Rhombohedral Trilayer Graphene}

Suppose that the BdG Hamiltonian describing superconducting rhombohedral trilayer graphene (R3G) is given by
\begin{equation}
    \mathcal{H}_{\mathrm{BdG}, \xi}(\mathbf{k}, \mathbf{q})
    =
    \left(
    \begin{array}{cc}
         \mathcal{H}_\xi\left(\mathbf{k} + \frac{\mathbf{q}}{2}\right)
         &
         \Delta_\xi(\mathbf{k})
         \\
         \Delta_\xi^\dagger(\mathbf{k})
         &
         - \mathcal{H}_{-\xi}^\mathrm{T}\left(-\mathbf{k} + \frac{\mathbf{q}}{2}\right)
    \end{array}
    \right),
    \label{BdG_Ham}
\end{equation}
in the Nambu basis
\begin{math}
    \Psi_{\xi, \mathbf{k}} = (c_{\xi, \mathbf{k} + \mathbf{q}/2}, c^\dagger_{-\xi, -\mathbf{k} + \mathbf{q}/2})^\mathrm{T}.
\end{math}
Here, $\mathbf{k}$ is the relative momentum about valley $\xi \mathbf{K}$, in which $\xi = \pm 1$, and $\mathbf{q}$ is the momentum of the Cooper pair.
The low-energy normal-state band structure is described by $\mathcal{H}_\xi(\mathbf{k})$, and $\Delta_\xi(\mathbf{k})$ is the pairing gap function.
We assume that pairing occurs between states near the $\mathbf{K}$ and $-\mathbf{K}$ Fermi surfaces.
For simplicity, we will take the pairing gap function to be momentum-independent and spin singlet,
\begin{equation}
    \Delta_{\xi}(\mathbf{k}) = \Delta_0 (i \sigma^y).
\end{equation}
Although the pairing gap matrix is taken to be spin singlet, spin-orbit coupling can induce spin triplet components in the anomalous pairing correlations~\cite{Gorkov2001, Zhou2016}.

The normal-state band Hamiltonian contains spin-independent intra- and interlayer hopping, proximity-induced spin-orbit coupling, and an exchange field describing the canted ferromagnetic order,
\begin{equation}
    \mathcal{H}_\xi(\mathbf{k}) = 
    -\mu\mathbbm{1}_{12 \times 12}
    +
    \mathcal{H}_{0, \xi}(\mathbf{k})
    +
    \mathcal{H}_{\mathrm{SOC}, \xi}(\mathbf{k})
    +
    \mathcal{H}_\mathrm{ex}.
    \label{SMeq:R3G_Ham}
\end{equation}
In the following, we focus on rhombohedral trilayer graphene as a representative example.
The spin-independent Hamiltonian, in the basis of $(A_1, B_1, A_2, B_2, A_3, B_3)^\mathrm{T}$ is given by~\cite{Zhang2010}
\begin{equation}
    \mathcal{H}_{0, \xi}(\mathbf{k})
    =
    \left(
    \begin{array}{ccc}
        h_1(\mathbf{k})
        &
        h_{\mathrm{inter}, \xi}(\mathbf{k})
        &
        t_{13}
        \\
        h_{\mathrm{inter}, \xi}^\dagger(\mathbf{k})
        &
        h_2 (\mathbf{k})
        & 
        h_{\mathrm{inter}, \xi}(\mathbf{k})
        \\
        t_{13}^\dagger
        &
        h_{\mathrm{inter}, \xi}^\dagger(\mathbf{k})
        &
        h_3(\mathbf{k})
    \end{array}
    \right)\otimes \sigma^0,
\end{equation}
in which
\begin{subequations}
    \begin{align}
        h_1(\mathbf{k})
        &=
        \left(
        \begin{array}{cc}
             \Delta_1 + \Delta_2 + \delta
             &
             \hbar v_0 \pi_{\xi}^*
             \\
             \hbar v_0 \pi_{\xi}
             &
             \Delta_1 + \Delta_2
        \end{array}
        \right),
        \\
        h_2(\mathbf{k})
        &=
        \left(
        \begin{array}{cc}
             -2 \Delta_2
             &
             \hbar v_0 \pi_{\xi}^*
             \\
             \hbar v_0 \pi_{\xi}
             &
             -2 \Delta_2
        \end{array}
        \right),
        \\
        h_3 (\mathbf{k})
        &=
        \left(
        \begin{array}{cc}
             -\Delta_1 + \Delta_2
             &
             \hbar v_0 \pi_{\xi}^*
             \\
             \hbar v_0 \pi_{\xi}
             &
             - \Delta_1 + \Delta_2 + \delta
        \end{array}
        \right)
        \\
        t_{13}
        &=
        \left(
        \begin{array}{cc}
             0 & \frac{\gamma_2}{2}
             \\
             0 & 0
        \end{array}
        \right),
        \\
        h_{\mathrm{inter}, \xi}(\mathbf{k})
        &=
        \left(
        \begin{array}{cc}
            \hbar v_4 \pi_\xi^*
            &
            \hbar v_3 \pi_\xi
            \\
            \gamma_1
            &
            \hbar v_4 \pi^*_\xi
        \end{array}
        \right).
    \end{align}
\end{subequations}
Above, we define $\pi_\xi \equiv \xi k_x + ik_y$,
where $\mathbf{k}$ is relative to the valley center $\xi \mathbf{K}$, and $\sigma^0$ is the $2 \times 2$ identity matrix in spin-space.
Here, $\hbar v_0 = (\sqrt{3}/2)a_0\gamma_0$, where $\gamma_0$ denotes the intralayer nearest-neighbour hopping; 
$\gamma_1$ the dominant interlayer dimer hopping; $\gamma_2$ the direct hopping between outer nondimer sites; 
$\hbar v_3 = (\sqrt{3}/2)a_0\gamma_3$ and $\hbar v_4 = (\sqrt{3}/2)a_0\gamma_4$, with $\gamma_3$ and $\gamma_4$ corresponding to skew interlayer hopping responsible for trigonal warping and particle-hole asymmetry, respectively;
$\delta$ the sublattice/dimer on-site energy offset; 
$\Delta_1$ the electrostatic potential asymmetry from the displacement field; 
and $\Delta_2$ the middle-layer electrostatic offset.
The strong interlayer hopping $\gamma_1$ hybridizes $B_1$--$A_2$ and $B_2$--$A_3$ into dimerized states.
Consequently, the low-energy electronic states reside predominantly at non-dimer sites $A_1$ and $B_3$.

The proximity-induced spin-orbit coupling consists of proximitized Ising and Rashba terms localized on the layer adjacent to the substrate,
\begin{equation}
    \mathcal{H}_{\mathrm{SOC}, \xi} (\mathbf{k})
    =
    \left(
    \begin{array}{ccc}
        \mathcal{H}_{I, \xi} + \mathcal{H}_{R, \xi}
        & 0 & 0
        \\
        0 & 0 & 0
        \\
        0 & 0 & 0
    \end{array}
    \right),
\end{equation}
where each block acts on combined sublattice and spin space.
Above, we take the first layer to be adjacent to the TMD.
The Ising spin-orbit coupling is diagonal in sublattice space and is given by
\begin{equation}
    \mathcal{H}_{I, \xi}
    =
    \frac{\Delta_I}{2} \xi \tau^0 \otimes \sigma^z ,
\end{equation}
in which $\sigma^\mu$ and $\tau^\nu$ are Pauli matrices in spin and sublattice space, respectively.
In contrast, the Rashba spin-orbit coupling is off-diagonal in sublattice space,
\begin{equation}
    \mathcal{H}_{R, \xi}
    =
\Delta_R
    (\xi \tau^x \otimes \sigma^y
    -
    \tau^y \otimes \sigma^x).
\end{equation}
Lastly, the canted ferromagnetic order is described by
\begin{equation}
    \mathcal{H}_\mathrm{ex}
    = 
    -J_\mathrm{ex} \mathbbm{1}_{3\times3} \otimes \tau^0 \otimes \mathbf{m} \cdot \boldsymbol{\sigma},
\end{equation}
in which $J_\mathrm{ex}$ is the strength of the exchange coupling.

\subsection{Magnetoelectric Coupling}

We next calculate the mixed magnetoelectric coupling, described by the following Lifshitz invariant,
\begin{equation}
    f_\mathrm{ME}
    = 
    \sum_{i, j}
q_i \Lambda_{q_i, m_j} m_j. \end{equation}
We expand the BdG Hamiltonian to linear order in $\mathbf{q}$ and $\mathbf{m}$,
\begin{equation}
    \mathcal{H}_{\mathrm{BdG}, \xi}(\mathbf{k}; \mathbf{q}, \mathbf{m})
    =
    \mathcal{H}^{(0)}_{\mathrm{BdG}, \xi}(\mathbf{k})
    +
    q_i \gamma_{\xi, q_i} (\mathbf{k})
    +
    m_j \gamma_{\xi, m_j}
    +
    \cdots,
\end{equation}
in which
\begin{equation}
    \mathcal{H}^{(0)}_{\mathrm{BdG}, \xi}(\mathbf{k})
    =
\mathcal{H}_{\mathrm{BdG}, \xi}(\mathbf{k}; \mathbf{q} =0, \mathbf{m} = 0)
\end{equation}
The unperturbed Green's function is given by
\begin{equation}
    \mathcal{G}_\xi^{(0)}(\mathbf{k}; i\omega_n)
    =
    (i\omega_n - \mathcal{H}_{\mathrm{BdG}, \xi}^{(0)}(\mathbf{k}))^{-1},
\end{equation}
and the vertices are
\begin{align}
        \gamma_{\xi, q_i} (\mathbf{k})
        =
        \left.
        \frac{\partial \mathcal{H}_{\mathrm{BdG}, \xi}(\mathbf{k}; \mathbf{q}, \mathbf{m})}{\partial q_i}
        \right|_{\mathbf{q} = \mathbf{m} = 0},
        \hspace{2em}
        \gamma_{\xi, m_j}
        =
        \left.
        \frac{\partial \mathcal{H}_{\mathrm{BdG}, \xi}(\mathbf{k}; \mathbf{q}, \mathbf{m})}{\partial m_j}
        \right|_{\mathbf{q} = \mathbf{m} = 0}.
    \end{align}
The thermodynamic potential is given by
\begin{equation}
    \Omega = - \frac{T}{2} \sum_{\xi, \omega_n}
    \int \frac{\mathrm{d}^2 k}{(2\pi)^2}
    \mathrm{Tr} \ln [-\mathcal{G}_\xi^{-1}(\mathbf{k}; i\omega_n)].
\end{equation}

For the given convention in Eq.~\eqref{BdG_Ham}, the phase-gradient vertex is
\begin{equation}
    \gamma_{\xi, q_i}(\mathbf{k})
    =
    \frac{1}{2}
    \left(
    \begin{array}{cc}
        \left.\partial_{k_i'} \mathcal{H}_\xi(\mathbf{k}')\right|_{\mathbf{k}' = \mathbf{k}}& 0
        \\
        0
        &
        \left. - \partial_{k_i'} \mathcal{H}_{-\xi}^\mathrm{T}(\mathbf{k}')\right|_{\mathbf{k}' = -\mathbf{k}}
    \end{array}
    \right),
\end{equation}
and the magnetic vertex is
\begin{equation}
    \gamma_{\xi, m_j} =
    \left(
    \begin{array}{cc}
        -J_\mathrm{ex} \mathbbm{1}_{3\times3} \otimes \tau^0 \otimes \sigma^j
        & 0
        \\
        0 &
        J_\mathrm{ex} (\mathbbm{1}_{3\times3} \otimes \tau^0 \otimes \sigma^j)^\mathrm{T}
    \end{array}
    \right).
\end{equation}

Expanding the free energy to second order gives the mixed response,
\begin{equation}
    \Lambda_{q_i, m_j}
    =
    \frac{T}{2}
    \sum_{\xi = \pm}
    \sum_{\omega_n}
    \int
    \frac{\mathrm{d}^2 k}{(2\pi)^2}
    \mathrm{Tr}
    \Big[
    \mathcal{G}_\xi^{(0)}(\mathbf{k}; i\omega_n)
    \gamma_{\xi, q_i}(\mathbf{k})
    \mathcal{G}_\xi^{(0)}(\mathbf{k}; i\omega_n)
    \gamma_{\xi, m_j}
    \Big].
    \label{one_loop.susceptibility}
\end{equation}
A finite $\Lambda_{q_i, m_j}$ for in-plane magnetization $j \in \{x,y\}$ requires in-plane spin-momentum locking.
In the present model, this is supplied by the Rashba spin-orbit coupling, while Ising spin-orbit coupling alone is insufficient.
It should be noted that an analogous derivation of Lifshitz invariants appears in the work by Mineev and Samokhin~\cite{Mineev2008} for multicomponent superconducting order parameters.

It is convenient to perform the Matsubara summation in the BdG eigenbasis $\{|u_{\xi, n} (\mathbf{k}) \rangle\}$, satisfying
\begin{equation}
    \mathcal{H}_{\mathrm{BdG}, \xi}^{(0)}(\mathbf{k})
    |u_{\xi, n} (\mathbf{k}) \rangle
    =
    E_{\xi, n}(\mathbf{k})
    |u_{\xi, n} (\mathbf{k}) \rangle.
\end{equation}
The one-loop susceptibility in Eq.~\eqref{one_loop.susceptibility} can be expressed as
\begin{equation}
    \Lambda_{q_i, m_j}
    =
    \frac{1}{2}
    \sum_{\xi = \pm}
    \sum_{n, m}
    \int \frac{\mathrm{d}^2 k}{(2\pi)^2}
    \frac{n_F(E_{\xi, n}) - n_F(E_{\xi, m})}{E_{\xi, n} - E_{\xi, m}}
    \langle u_{\xi, n} | \gamma_{\xi, q_i} | u_{\xi, m} \rangle
    \langle u_{\xi, m} | \gamma_{\xi, m_j} | u_{\xi, n} \rangle,
\end{equation}
in which $n_F(E) = (1 + e^{E/T})^{-1}$ is the Fermi function.
Above, we use the spectral representation,
\begin{equation}
    \mathcal{G}_\xi^{(0)} (\mathbf{k}; i\omega_n)
    =
    \sum_n
    \frac{ |u_{\xi, n}(\mathbf{k})\rangle \langle u_{\xi, n}(\mathbf{k})|}{i\omega_n - E_{\xi, n}(\mathbf{k})},
\end{equation}
and the following Matsubara sum for nondegenerate states,
\begin{gather}
    T
    \sum_{\omega_n}
    \frac{1}{(i\omega_n - E_\alpha) (i\omega_n - E_\beta)}
=
    \frac{n_F(E_\alpha) - n_F(E_\beta)}{E_\alpha - E_\beta}.
\end{gather}
For degenerate states, the weighting factor should be replaced by
\begin{equation}
    \lim_{E_n \rightarrow E_m}
    \frac{n_F(E_n) - n_F(E_m)}{E_n - E_m}
    =
    \left.\frac{\mathrm{d} n_F(E)}{\mathrm{d} E}\right|_{E_n}
=
    - \frac{1}{4T} \mathrm{sech}^2
    \left(
    \frac{E_n}{2T}
    \right).
\end{equation}
The one-loop susceptibility is given by
\begin{equation}
    \Lambda_{q_i, m_j}
    =
    \frac{1}{2}
    \sum_{\xi = \pm}
    \sum_{n, m}
    \int \frac{\mathrm{d}^2 k}{(2\pi)^2}
    W_{nm}^\xi
    \langle u_{\xi, n} | \gamma_{\xi, q_i} | u_{\xi, m} \rangle
    \langle u_{\xi, m} | \gamma_{\xi, m_j} | u_{\xi, n} \rangle,
\end{equation}
in which
\begin{equation}
    W_{nm}^{\xi}
    =
    \begin{cases}
        \frac{n_F(E_{\xi, n}) - n_F(E_{\xi, m})}{E_{\xi, n} - E_{\xi, m}}
        & E_{\xi, n} \neq E_{\xi, m}
        \\
        - \frac{1}{4T} \mathrm{sech}^2
        \left(
        \frac{E_{\xi, n}}{2T}
        \right)
        &
        E_{\xi, n} = E_{\xi, m}
    \end{cases}
\end{equation}
Lastly, for Rashba spin-orbit coupling, it follows that
\begin{equation}
    \Lambda_{q_x, m_y} = - \Lambda_{q_y, m_x} \equiv \Lambda_\mathrm{ME}.
\end{equation}

\subsection{Superfluid Stiffness}

We next detail the calculation of the superfluid stiffness tensor, $\rho_s^{ij}$, used to determine the magnetoelectric coupling $\lambda$.
We retain the conventional intraband contribution from a chosen active band $n$, with dispersion $\varepsilon_n(\mathbf{k})$ and chemical potential $\mu$.
The corresponding group velocity is given by
\begin{equation}
    v_{n, i}(\mathbf{k}) = 
    \frac{1}{\hbar}
    \frac{\partial \varepsilon_n(\mathbf{k})}{\partial k_i},
\end{equation}
in which $i = x, y$.
For simplicity, we consider a uniform $s$-wave superconducting pairing gap with amplitude $\Delta_0$.
The Bogoliubov quasiparticle energy is then given by
\begin{equation}
    E_n(\mathbf{k})
    =
    \sqrt{(\varepsilon_n(\mathbf{k})-\mu)^2 + \Delta_0^2}.
\end{equation}
At zero temperature, the conventional phase stiffness is given by
\begin{equation}
    \rho_s^{ij} 
    =
    \frac{1}{4}
    \int
    \frac{\mathrm{d}^2 k}{(2\pi)^2}
    \frac{\Delta_0^2}{E_n^3}
    \frac{\partial \varepsilon_n (\mathbf{k})}{\partial k_i}
    \frac{\partial \varepsilon_n (\mathbf{k})}{\partial k_j}
    = 
    \frac{\hbar^2}{4}
    \int \frac{\mathrm{d}^2 k}{(2\pi)^2}
    \frac{\Delta_0^2}{E_n^3(\mathbf{k})}
    v_{n, i} (\mathbf{k})
    v_{n, j} (\mathbf{k}),
    \label{SMEq:SF_stiffness}
\end{equation}
and should be evaluated for both valleys.
The above superfluid stiffness does not include the quantum geometric contribution~\cite{Peotta2015, Toermae2022}.
The quantum geometric contribution is small relative to the contribution in Eq.~\eqref{SMEq:SF_stiffness} by a factor of the order of $|\Delta_0|^2/E_F^2$, where $E_F$ is the Fermi energy, up to logarithmic corrections. A rough estimate of this factor for R3G gives that the geometric contribution is about $\sim 10^{-2}$ to $10^{-4}$ of the conventional contribution in Eq.~\eqref{SMEq:SF_stiffness}.

\subsection{Values of Magnetoelectric Coupling}

The magnetoelectric coupling $\lambda = \Lambda_\mathrm{ME}/\rho_s$ is determined by both the Lifshitz coefficient $\Lambda_\mathrm{ME}$ and superfluid stiffness $\rho_s$.
For the Hamiltonian in Eq.~\eqref{SMeq:R3G_Ham}, we consider the following tight-binding parameters~\cite{Zhang2010, Koh2024},
\begin{equation}
\begin{alignedat}{3}
\gamma_0 &= 3100\,\text{meV},
&\qquad \gamma_1 &= 380\,\text{meV},
&\qquad \gamma_2 &= -15\,\text{meV},
\\
\gamma_3 &= -290\,\text{meV},
&\qquad \gamma_4 &= -141\,\text{meV},
&\qquad \delta &= -10.5\,\text{meV},
\\
\Delta_2 &= -2.3\,\text{meV},
&\qquad \Delta_1 &= 5\,\text{meV},
&\qquad a_0 &= 0.246\,\text{nm}.
\end{alignedat}
\end{equation}
The dependence of $\lambda$ on the spin-orbit and exchange couplings is primarily inherited from $\Lambda_\mathrm{ME}$, as the superfluid stiffness is comparatively insensitive to these parameters over the range considered.
At weak coupling, we find that $\Lambda_\mathrm{ME}$ scales approximately linearly with the exchange coupling and Rashba spin-orbit coupling, $\Lambda_\mathrm{ME} \propto J_\mathrm{ex} \Delta_R$.

As discussed in the main text, the reported values of Rashba spin-orbit coupling vary widely.
For conservative values of Rashba spin-orbit coupling on order of $0.1$--$1$ meV and bare superfluid stiffnesses on order of a few meV, we obtain
$|\lambda m_0| \sim 10^{-6}$--$10^{-5}$ nm$^{-1}$, corresponding to pitches of order $\ell_0 \sim 0.1$--$1$ mm.
Moreover, although Eq.~\eqref{SMEq:SF_stiffness} yields a value of superfluid stiffness on the order of a few meV, 
recent local magnetic measurements of R3G/WSe$_2$ report a spatially varying stiffness reaching $\rho_s \simeq 2.3$ K, or approximately $0.2$ meV, at $T = 70$ mK~\cite{Zhang2026}.
This measured stiffness is substantially smaller than the estimate based on the bare band structure.
For stronger Rashba spin-orbit coupling together with a reduced superfluid stiffness, the coupling can increase to $|\lambda m_0| \sim 10^{-4}$--$10^{-3}$ nm$^{-1}$, giving characteristic pitches from a few to tens of microns.

Notably, a large microscopic magnetoelectric coupling is not required for the smectic-like regime.
In absence of pinning of the transverse magnetic fluctuations, any finite $\lambda$ causes the long-wavelength transverse superfluid stiffness to vanish.
The helical pitch $\ell_0$ provides a useful diagnostic of the magnitude of the magnetoelectric coupling, but does not by itself determine the crossover scale at which the smectic-like elasticity becomes relevant.
Rather, this scale also depends on the magnetic and superfluid stiffnesses, as discussed in the main text and S.M. Sec.~\ref{SM:subsec:length_scales}.
Thus, even weak microscopic magnetoelectric coupling can qualitatively modify the asymptotic phase response and finite-temperature phase diagram.
For sufficiently weak magnetoelectric coupling, however, the crossover length can exceed the dimensions of a finite system, making the asymptotic smectic-like regime difficult to experimentally access.

 \section{Sixfold Anisotropy and Pinning of Ferromagnetic Order}

\label{appendix:pinning}

In this section, we detail the pinning of the ferromagnetic order from the sixfold anisotropy.
In the absence of an in-plane exchange field, the system has $C_3$ and time-reversal symmetry prior to magnetic condensation.
After the magnitude of the spin polarization condenses, the remaining degree of freedom is the orientation angle.
The leading anisotropy allowed by $C_3$ and time reversal is a sixfold clock anisotropy.

We consider the free energy of the spin polarization,
\begin{equation}
    F_m
    =
    \int \mathrm{d}^2 r 
    \left(
    \frac{K_m}{2} (\partial_i \mathbf{m})^2
    -
    \mathbf{h}_\parallel \cdot \mathbf{m}
\right),
\end{equation}
where we omit amplitude terms $\frac{r_m}{2}\mathbf{m}^2 + \frac{u_m}{4} (\mathbf{m}^2)^2$ to focus on the orientation angle.
A quadratic term such as $m_y^2$ is not allowed by $C_3$ symmetry alone but should instead be viewed as an effective pinning term generated by (i)~expanding locally about one of the six anisotropy-selected orientations, (ii)~an in-plane exchange field, or (iii)~additional symmetry breaking such as strain.

\subsection{Sixfold Anisotropy}

We begin by examining the role of the sixfold anisotropy in absence of an exchange field.
In the following, we use a fixed-amplitude approximation and parameterize the in-plane spin polarization as
\begin{equation}
    \mathbf{m}(\mathbf{r}) = m_0 (\cos \varphi(\mathbf{r}), \sin \varphi(\mathbf{r}))^\mathrm{T}.
\end{equation}
The free energy describing the in-plane magnetization with sixfold anisotropy is given by
\begin{equation}
    F_\varphi
    =
    \int \mathrm{d}^2 r
    \left(
    \frac{\kappa}{2} (\boldsymbol{\nabla} \varphi)^2 - g \cos (6 \varphi)
    \right),
\end{equation}
in which $\kappa = K_m m_0^2$ and $g>0$ is the bare sixfold clock anisotropy.
For $g=0$, the Gaussian free-field correlator is given by
\begin{equation}
    \langle \varphi(\mathbf{r}) \varphi(0) \rangle = - \frac{T}{2\pi \kappa} \ln \frac{r}{a} + \mathrm{const},
\end{equation}
in which $a$ is a UV cutoff.
It follows that
\begin{align}
    \langle e^{i p \varphi(\mathbf{r})} e^{-i p \varphi(0)} \rangle
    \propto
    \exp
    \left(
    - \frac{p^2}{2} \frac{T}{\pi \kappa} \ln \frac{r}{a}
    \right)
    =
    \left(
    \frac{a}{r}
    \right)
    ^{p^2 T/(2\pi \kappa)},
\end{align}
which yields scaling dimension
\begin{equation}
    d_p = \frac{p^2 T}{4 \pi \kappa},
\end{equation}
for $\langle e^{ip\varphi(\mathbf{r})} e^{-i p \varphi(0)}\rangle \sim r^{-2 d_p}$.
For $p = 6$, the sixfold anisotropy has scaling dimension
\begin{equation}
    d_6 = \frac{9T}{\pi \kappa}.
\end{equation}
Close to the lower transition, the RG equations take the BKT form~\cite{Jose1977},
\begin{equation}
    \frac{\mathrm{d} y}{\mathrm{d} \ell} = xy,
    \hspace{2em}
    \frac{\mathrm{d} x}{\mathrm{d} \ell} = Ay^2,
\end{equation}
in which $A > 0$, $x = 2 - d_6$, and $y$ is the dimensionless sixfold-anisotropy coupling.
As such, the sixfold anisotropy is relevant for $d_6 < 2$, or equivalently, for
\begin{equation}
\frac{\kappa_R(T)}{T}
    > \frac{9}{2\pi},
\end{equation}
where $\kappa_R$ is the renormalized angular stiffness.
The lower transition into the $Z_6$-ordered phase is given by
\begin{equation}
    \frac{\kappa_R(T_6)}{T_6} = \frac{9}{2\pi}.
\end{equation}
Neglecting stiffness renormalization, one may approximate $\kappa_R \approx K_m m_0^2$.
For $T < T_6$, the sixfold anisotropy flows to strong coupling and locks the in-plane angle $\varphi$ to one of six preferred orientations.

Close to $T_6$, the RG scale at which the anisotropy becomes strong is given by
\begin{equation}
    \ell_* \sim \frac{b}{\sqrt{(T_6 - T)/T_6}},
\end{equation}
in which $b$ is a phenomenological coefficient of order unity.
This defines the characteristic length scale,
\begin{equation}
    \xi_6 = a e^{\ell_*} \sim a \exp
    \left(
    \frac{b}{\sqrt{(T_6 - T)/T_6}}
    \right).
    \label{xi_6_scale}
\end{equation}
Once the sixfold anisotropy locks the angle, the angular mode acquires a mass term.
Taking the locked direction to be $\varphi = 0$, the free energy is given by
\begin{equation}
    F_\varphi = \int \mathrm{d}^2 r \left(
    \frac{\kappa}{2} (\boldsymbol{\nabla} \varphi)^2 + \frac{\mu_\perp^{(6)} m_0^2}{2} \varphi^2
    \right),
\end{equation}
in which $\mu_\perp^{(6)} > 0$ is the effective transverse pinning strength from the sixfold anisotropy.
The associated characteristic length is 
\begin{math}
    \xi_6^2 = {\kappa}/{(\mu_\perp^{(6)} m_0^2)}.
\end{math}
From the characteristic scale in Eq.~\eqref{xi_6_scale}, we obtain the estimate for the pinning strength,
\begin{equation}
    \mu_\perp^{(6)} 
    \sim
    \frac{\kappa}{m_0^2 \xi_6^2}
    \sim
    \frac{K_m}{a^2}
    \exp
    \left(
    - \frac{2b}{\sqrt{(T_6 - T)/T_6}}
    \right).
\end{equation}
This is the renormalized low-energy pinning scale.
At the bare level, expanding the sixfold term gives
\begin{equation}
    -g \cos (6\varphi) \approx -g + 18 g \varphi^2.
\end{equation}
Comparing with the above mass term, we find
\begin{equation}
    \mu_{\perp, \mathrm{bare}}^{(6)} m_0^2 = 36g.
\end{equation}
Near the transition into the $Z_6$-ordered phase, however, the renormalized estimate above is generally more useful.

\subsection{Effects of an In-Plane Exchange Field}

An in-plane exchange field explicitly selects a direction for $\mathbf{m}$ and can give an additional contribution to the transverse mass.
Moreover, the in-plane exchange field explicitly breaks the residual $Z_6$ symmetry.
Consequently, the lower transition is rounded into a crossover, while the transverse mode remains massive.

Suppose the in-plane exchange field is along the $x$ direction, with $\mathbf{h}_\parallel = h_\parallel \hat{\mathbf{x}}$.
For the fixed-amplitude approximation, the contribution entering the free energy is given by
\begin{equation}
    - \mathbf{h}_\parallel \cdot \mathbf{m} = - h_\parallel m_0 \cos \varphi \approx - h_\parallel m_0  + \frac{h_\parallel m_0}{2} \varphi^2.
\end{equation}
Comparing to the term $\frac{1}{2} \mu_\perp m_y^2 \approx \frac{1}{2}\mu_\perp m_0^2 \varphi^2$, this gives a transverse mass term,
\begin{equation}
    \mu_\perp^{(h)} = \frac{h_\parallel}{m_0}.
\end{equation}
When the exchange field is aligned with the preferred locked direction, the total transverse pinning strength is approximately
\begin{equation}
    \mu_\perp^\mathrm{eff} = \mu_\perp^{(6)} + \mu_\perp^{(h)} = \mu_\perp^{(6)} + \frac{h_\parallel}{m_0}.
\end{equation}
This effective mass is the quantity that cuts off the smectic regime and restores a finite long-distance transverse phase stiffness.